\documentclass[runningheads]{llncs}

\usepackage[T1]{fontenc}
\usepackage{graphicx}

\usepackage{graphicx}
\usepackage{longtable}
\usepackage{amsfonts}
\usepackage{amssymb}
\usepackage{bm}
\usepackage[utf8]{inputenc}
\usepackage[english]{babel}
\usepackage{xstring}
\usepackage{setspace}
\usepackage{array}
\usepackage{subfig}
\usepackage{enumitem}
\usepackage{multirow}
\usepackage{algorithmic}
\usepackage{dutchcal} 			
\usepackage{ctable} 				
\usepackage{comment}
\usepackage[ruled,vlined,linesnumbered]{algorithm2e}
\usepackage{amsmath,amsfonts,amssymb}
\usepackage{calligra}
\usepackage{balance}

\newcommand\modes {\sf md}

\newcommand\True {\textsf{t}}
\newcommand\False {\textsf{f}}
\newcommand{\X}[1] {#1} 					
\newcommand{\V}[1] {\textsf{#1}} 			
\newcommand{\VA}[1] {\textsf{#1}'} 			
\newcommand{\VAD}[1] {\textsf{#1}^d{'}}	
\newcommand{\VR}[1] {\textsf{#1}} 			
\newcommand{\VRD}[1] {\textsf{#1}^d}		
\newcommand{\XF}[1] {\Hat{#1}} 			
\newcommand{\XFD}[1] {\Hat{#1^d}} 		
\newcommand{\VF}[1] {{\cal{#1}}}			

\newcommand\Defs {{:=}\,}
\newcommand{\NF}{\mathbin\sim} 			
\newcommand{\MD}{{\modes_{\cal P}}}		
\newcommand{\Sign}[1] {\tilde{s}(#1)}		
\newcommand{\Bold}[1] {\bm{#1}} 			

\newcommand{\Sensor}[1]{
    \IfEqCase{#1}{
       {IW512}{L1}
       {IW576}{L2}
       {IW544}{L3}
       {IW554}{P1}
       {IW560}{P2}
        {#1}{#1}
    }
}

\newcommand{\Imply}{\Rightarrow}

\newcommand{\TE}[1]{{\scriptsize\sffamily #1}} 
\newcolumntype{M}[1]{>{\centering\arraybackslash}m{#1}}
\newcolumntype{N}{@{}m{0pt}@{}}

\newcommand{\Eq}[1]{(\ref{#1})}
\newcommand{\Def}[1]{Def. \ref{#1}}
\newcommand{\Fig}[1]{Fig. \ref{#1}}
\newcommand{\Table}[1]{Table \ref{#1}}
\newcommand{\Sec}[1]{Section \ref{#1}}

\newcommand{\Theorem}[1]{Theorem \ref{#1}}

\newcommand{\Assump}[1]{Assumption \ref{#1}}
\newcommand{\Corol}[1]{Corollary \ref{#1}}

\newtheorem{assumption}{Assumption}
\newtheorem{axiom}{Axiom}

\begin{document}

\title{From Function to Failure}

\author{Hamid Jahanian}

\institute{Macquarie University, Sydney, Australia\\
\email{hamid.jahanian@hdr.mq.edu.au}}

\maketitle             

\begin{abstract}
Failure Mode Reasoning (FMR) is a method for formal analysis of system-related faults. The method was originally developed for identifying failure modes of safety-critical systems based on an analysis of their programs. In this paper, we generalize the method and present a mathematical framework for its use in model-based system and safety analyses. We explain the concepts, formalize the method, formulate models for example systems, and discuss the practical application of the method.

\keywords{Failure Mode Reasoning  \and FMR.}
\end{abstract}

\section{Introduction}\label{Sec_Intro}

The term `failure' is defined as the state of \emph{not} functioning.\footnote{Failure is the termination of the ability of a functional unit to provide a required function or operation of a functional unit in any way other than as required \cite{Ref_186}.} This, however, does not mean that we can somehow negate the definition of a system function to derive its failure model. Functional and failure models are of two different natures and dealt with in two different spaces.

Failure Mode Reasoning (FMR) was an attempt to bridge the two spaces. By implementing an automated reasoning process, FMR derives the failure model of a system from the definition of its functional behavior. The method was originally designed for identifying failure modes of Safety Instrumented Systems (SIS)\footnote{In the process industry, SIS is a system that performs safety functions to protect the plant from major hazards \cite{Ref_190}.} based on an analysis of their software program \cite{Ref_200_0,Ref_225_0,Ref_230_0,Ref_295_0}. However, the application of FMR is not limited to SIS programs; any system comprising interrelated components, where the component functions and the interactions between them are defined, can be modeled and analyzed in FMR. 

Various other methods exist for system failure analysis, including the conventional methods, such as Fault Tree Analysis (FTA) \cite{Ref_176,Ref_168} and Failure Mode and Effect Analysis (FMEA) \cite{Ref_181}, and the more advanced methods, such as HiP-HOPS \cite{Ref_140,Ref_71,Ref_139} and Model-Based Safety Analysis (MBSA) \cite{Ref_306,Ref_307}. FMR differs from the majority of these methods in two main aspects: a) FMR does not require a separate failure model and it extracts it directly from the function model; and b) FMR does not rely on analyst's perception of failure behavior because it uses an automated process to create the model. These two qualities are particularly important in complex systems, for which by-hand analysis of interrelations between numerous components is not feasible.

In this paper, we introduce FMR in its general form and elaborate on its mathematical principles. Our goal is to provide a framework for applying FMR in generic model-based system and safety analysis. The rest of this paper is organized as follows: we begin with the basic definitions and assumptions used in FMR in \Sec{Sec_Basics}. We then explain the FMR analysis process, including abstraction, causation, and composition, in \Sec{Sec_FMR}. We use a small worked example for demonstration in Sections \ref{Sec_Basics} and \ref{Sec_FMR}. In \Sec{Sec_CompFunctions}, we expand the discussion by providing some general modeling rules and deriving FMR models for common system structures DNF, CNF, and KooN.\footnote{DNF (or CNF): Disjunctive (or Conjunctive) Normal Form; KooN: K-out-of-N.} In \Sec{Sec_Discussion}, we discuss three special topics related to FMR: modeling function loops, measuring the impact of parametric changes, and the application of FMR. In \Sec{Sec_RelatedWorks}, we cover a comparison between FMR and other relevant methods. 

\section{The context}\label{Sec_Basics}

A system is a combination of interacting components organized to achieve a purpose \cite{Ref_347}. Thus, a system can be modeled by a set of interconnected functions, where each individual function models the functional behavior of a corresponding component, and the interconnections between the functions represent the interactions between the components. 

\begin{figure}[!h]
\includegraphics[scale=0.65]{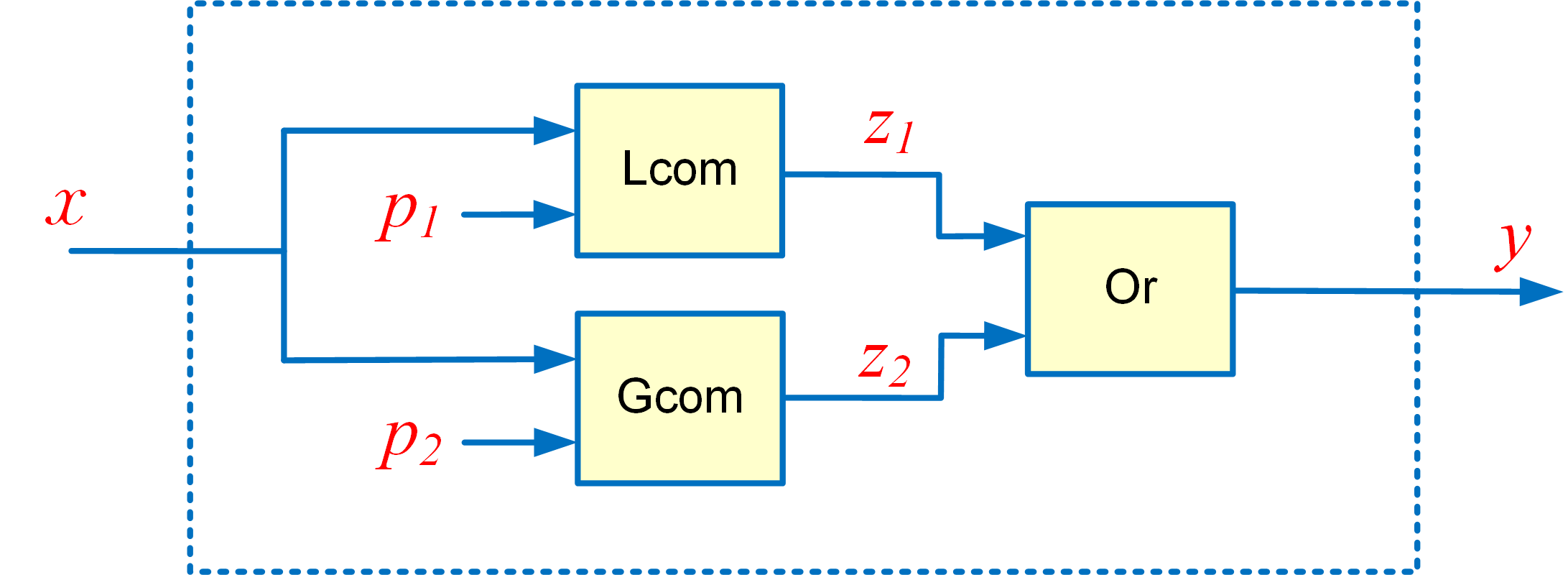}
\centering
\caption{A small system comprising three components}
\label{Fig_System1}
\end{figure}

For a small example, consider a system that is meant to read a real-valued input and decide whether the input value is outside a permissible range. \Fig{Fig_System1} shows one possible representation for this system. The model comprises three component functions: $Lcom$ for less comparison; $Gcom$ for greater comparison, and $Or$ for Boolean disjunction. The system output will be $\X{y}=\True$ if $\X{x}<\X{p}_1$ or $\X{x}>\X{p}_2$, and will be $\X{y}=\False$ if $\X{p}_1 \leq \X{x} \leq \X{p}_2$.\footnote{We use $\True$ and $\False$ for Boolean \emph{True} and \emph{False}.} 

In FMR, component functions are defined in mathematical terms. So the resultant system function will be a mathematical function too, except that a system function is (typically) much more complex than its constituting component functions. 

\begin{definition}\label{Def_Sys_Function}
Let ${\cal V}$ be a state space. A \emph{system} (or \emph{component}) \emph{function} is a function of type $f: {\cal V}\rightarrow {\cal V}$ that defines the functional behavior of a system (or component).
\end{definition}

The result (output) of function $f$ depends on its arguments (inputs), the architecture of $f$ itself, and, possibly, a number of `parameters.' A parameter can be a control setting that is adjusted by system designers to influence the behavior of the component (e.g., $\X{p}_1$ and $\X{p}_2$ in \Fig{Fig_System1}) or a modeling parameter that represents an aspect of a component (e.g., the travel time of a valve). We express function $f$ as follows:
\begin{align}
&\Bold{\X{y}}=f(\Bold{\X{x}}, \Bold{\X{p}}) \label{Eq_F}
\end{align}
where $\Bold{\X{y}}, \Bold{\X{x}}, \Bold{\X{p}} \in {\cal V}$ are the outputs, inputs, and parameters, respectively.\footnote{We use bold font for vectors and matrices. Also, note that we only separate between $\Bold{\X{x}}$ and $\Bold{\X{p}}$ for better clarity in our failure discussions. Mathematically speaking, $\Bold{\X{x}}$ and $\Bold{\X{p}}$ together represent one vector of arguments for $f$.} 

We said earlier that failure is the state of not functioning. An advantage of defining system functions in mathematical terms is that we can define their failure in numerical form too. 

\begin{definition}\label{Def_Sys_Failure}
Given the system function $\Bold{\X{y}}=f(\Bold{\X{x}}, \Bold{\X{p}})$, a \emph{failure} at output $\Bold{\X{y}}$ occurs when the `reported' value at $\Bold{\X{y}}$ differs from the `intended' value at $\Bold{\X{y}}$.
\end{definition}

Failure is the characteristic of a system output. For a more generic expression of deviation between the reported and intended values at an arbitrary system variable, we use the term `fault.'   

\begin{definition}\label{Def_Sys_Fault}
Let $\X{x}$ be a system variable, $\VR{x}$ the reported value at $\X{x}$ and $\VA{x}$ the intended value at $\X{x}$. We say $\X{x}$ is \emph{faulty} if $\VR{x} \neq \VA{x}$, and we say $\X{x}$ is \emph{fault-free} if $\VR{x} = \VA{x}$.
\end{definition}

In simple terms, given a system function $\Bold{\X{y}}=f(\Bold{\X{x}}, \Bold{\X{p}})$ and a failure at output $\Bold{\X{y}}$, FMR identifies the triggering faults at inputs $\Bold{\X{x}}$ and/or parameters $\Bold{\X{p}}$. A key assumption here is that the function model $f$ is correct. This is particularly important because the reasoning process of FMR is entirely based on the structure of this function. Thus, in FMR:

\begin{assumption}\label{Assump_1} 
The only possible causes for a system output failure are faulty inputs and parameters.
\end{assumption}

An output failure can present itself in the form of an `omission' or a `commission.' Omissions occur when the output of a system does \emph{not} change state when it is expected to, and commissions occur when the system output changes state when it is not expected to. This obviously implies the existence of an \emph{intended state} prior to the failure. We may as well say that omissions occur when the system output does not change to an intended state and commissions occur when the system output changes state to an unintended state. An obvious assumption in both cases is that the system is in an intended state prior to the occurrence of a failure. 

\begin{assumption}\label{Assump_2} 
The inputs and outputs of a system are in an intended state prior to a failure being observed at its output.
\end{assumption}

With Assumptions \ref{Assump_1} and \ref{Assump_2} together, we can be more specific about causes of an output failure:

\begin{assumption}\label{Assump_3}
An output failure can only occur following one of the following events:
\begin{enumerate}[label=(\roman*)]
\item an external fault occurs at the input and propagates through the system, or
\item some system parameters are faulty when an intentional change occurs at the input.
\end{enumerate}
\end{assumption}

Scenario (i) represents a situation in which the system is in normal operation and an external fault occurs at its input. Scenario (ii) shows a test setup, where intentional inputs are injected at the system input for validation testing, for instance. 

\section{The analysis process}\label{Sec_FMR}

FMR is a hierarchical analysis at three levels: variables, component functions, and system function. The method uses three techniques that correspond these hierarchical levels: \emph{abstraction} for variables, \emph{causation reasoning} for component functions, and logical \emph{composition} for system functions. In this section, we elaborate on these principles. For demonstration, we use the small system depicted in \Fig{Fig_System1}.\footnote{Readers can find more elaborate examples in \cite{Ref_200_0,Ref_225_0,Ref_295_0}.} 

\subsection{Abstraction and variables}\label{Sec_Abstraction}

In a general sense, abstraction is the process of mapping a representation of a problem onto a new representation, which helps solve the original problem by only preserving certain desirable properties \cite{Ref_296}. In FMR, variable faults are abstracted into `failure modes.'

\begin{definition}\label{Def_FM}
A \emph{failure mode} is a member of $\mathcal{P}=\{\VF{h}, \VF{l}, \VF{t}, \VF{f}, \VF{m}\}$ and it depicts the type of fault at a system variable. Let $\X{x}$ be a variable and $\VR{x}$ and $\VA{x}$ its reported and intended values, respectively. We denote the failure mode at $\X{x}$ by $\XF{x} \in \mathcal{P}$, where $\XF{x}$ is generated by $\MD(\VR{x}, \VA{x})$, as follows:
\begin{equation} \label{Eq_MD1}
\MD(\VR{x}, \VA{x}) \equiv \XF{x} ~\Defs~ 
\begin{cases}
\VF{h}, \text{ if $\VR{x}, \VA{x} \in \mathbb{R}$ and $\VR{x}>\VA{x}$} \\
\VF{l}, \text{ if $\VR{x}, \VA{x} \in \mathbb{R}$ and $\VR{x}<\VA{x}$} \\
\VF{t}, \text{ if $\VR{x}, \VA{x} \in \mathbb{B}$ and $\VR{x}=\True,~ \VA{x}=\False$} \\
\VF{f}, \text{ if $\VR{x}, \VA{x} \in \mathbb{B}$ and $\VR{x}=\False,~ \VA{x}=\True$} \\
\VF{m}, \text{ if $\VR{x}=\VA{x}$} 
\end{cases}
\end{equation}
\end{definition}

As an example, $\XF{x}=\VF{h}$ means that $\X{x}$ is reporting a \emph{higher} value than it should (i.e., the reported value at $\X{x}$ is \emph{higher} than the intended value). 

Note that $\VF{m}$ is not a fault, and we only define it for the completeness of $\MD$. We may also use $\VF{a}$ to indicate \emph{any} failure mode:\footnote{Logically, $\XF{x} = \VF{a}$ is an always-true sentence, and where it is used, it is only used for simplifying the expression.}
\begin{align}
& (\XF{x} = \VF{a}) \equiv (\XF{x}=\VF{h} \lor \XF{x}=\VF{m} \lor \XF{x}=\VF{l}), \text{ for $\X{x} \in \mathbb{R}$}. \label{Eq_MD5}\\
& (\XF{x} = \VF{a}) \equiv (\XF{x}=\VF{t} \lor \XF{x}=\VF{m} \lor \XF{x}=\VF{f}), \text{ for $\X{x} \in \mathbb{B}$}. \label{Eq_MD6}
\end{align}

By using failure mode abstraction, unnecessary numerical calculations are eliminated from the analysis and the problem is reduced to failure mode calculation. As an example, all the combinations of $\VR{x}, \VA{x} \in \mathbb{R}$ where $\VR{x} > \VA{x}$ are summarized into one failure mode $\XF{x}=\VF{h}$. What we need to do then is to compute with few simple failure modes instead of a wide range of numerical values.

Note that the abstraction used in FMR is not on concrete values or concrete deviations but rather on the deviation between a known, certain value and an unknown, probable value -- both of them related to the same system variable. Hence, FMR is a non-deterministic analysis, and failure modes represent uncertain events. 

\paragraph{Types of variables:} 
From the FMR perspective, a variable can be either `certain' or `suspicious.' Suspicious variables are those in which we search for faults (i.e., their current values may differ from their intended values). Certain variables, on the other hand, are those that are given (i.e., their current values are as intended and, thus, they cannot be faulty). 

Recall the $Gcom$ function from our example in \Fig{Fig_System1}: $\X{z}_2 = Gcom(\X{x}, \X{p}_2) \Defs (\X{x} > \X{p}_2)$. This function compares its input $\X{x}$ to a threshold parameter $\X{p}_2$ and produces $\X{z}_2=\True$ if $\X{x} > \X{p}_2$. In general, $\XF{z}_2=\VF{t}$ (i.e., when the expected value at $\X{z}_2$ is $\False$ but $\X{z}_2$ is reporting $\True$) implies that either $\XF{x}=\VF{h}$ or $\XF{p}_2=\VF{l}$. Now, if we know that the input has changed by intention (i.e., if $\X{x}$ is a `certain' variable) we can rule out $\XF{x} = \VF{h}$ and consider $\XF{p}_2 = \VF{l}$ as the only possible cause for $\XF{z}_2 = \VF{t}$.

\subsection{Causation and reasoning}\label{Sec_Causation}

Where a functional relationship between two variables $\X{x}$ and $\X{y}$ is known, we can also determine the causal relationship between their corresponding failure mode variables $\XF{x}$ and $\XF{y}$. Consider the Boolean negation function $\X{y} = Not(\X{x}) =\neg \X{x}$. From this functional definition, we can follow that a fault $\VF{t}$ (or $\VF{f}$) at input $\X{x}$ will cause an opposite fault $\VF{f}$ (or $\VF{t}$) at output $\X{y}$. If a failure at output $\X{y}$ is given, we can also reason in backward direction: $\XF{y}=\VF{t} \Imply \XF{x}=\VF{f} $ and $ \XF{y}=\VF{f} \Imply \XF{x}=\VF{t} $. 

Notice that these implication statements are only valid in the context of the original function $Not$. To include the function names in our failure models, we use the following form of triples, originally introduced in \cite{Ref_230_0}:
\begin{eqnarray}
\{ \XF{x}= \VF{f}\} &~~Not~~& \{\XF{y}=\VF{t}\} \\
\{ \XF{x}= \VF{t}\} &~~Not~~& \{\XF{y}=\VF{f}\}
\end{eqnarray}

In these triples, the left-hand side sentences indicate the cause of failure at the input of the function, and the sentences on the right show the effect of the failure at the output. 

In a rather general term, we refer to `$\{\phi_c\}~f~\{\phi_e \wedge \phi_p\}$' as a `failure scenario,' in which $\phi_c$ is the conclusion we can make about input failure modes based on $\phi_e \wedge \phi_p$ as a premise. The latter part itself comprises $\phi_e$ as the output failure mode and $\phi_p$ as any additional context information given. Hence, this expression can be read as follows: given the premise $\phi_p$, if $\phi_e$ occurs at the output of $f$, the cause must be the occurrence of $\phi_c$ at the input of $f$. 

Regardless of depictions and symbols, FMR is a logical reasoning process. Reasoning is the process of developing certainty on what we know less about, based on the certainty of what we know more about. Reasoning, in general, may be divided into deductive, inductive, and abductive forms \cite{Ref_278}. FMR employs deductive reasoning. The main question we try to answer in FMR is: Given what we know about a failure at the output of a known function, what can we say about faults at its input? The minimum information required to begin an FMR analysis is a given failure mode at the output. In some scenarios, additional information about the current values of variables may be given, which can help narrow down the scope of the search for faulty variables. 

From this perspective, we classify the types of reasoning into two main categories: 
\begin{enumerate}[label=(\roman*)]
\item value-independent, where we only use the failure modes defined in \Eq{Eq_MD1}.
\item value-dependent, where we use failure modes as well as some additional information related to the current \emph{values} of some of the variables.
\end{enumerate}

Selecting the right type of reasoning obviously depends on the information available prior to the analysis. Nonetheless, the stronger the prior knowledge, the more precise the final results. Take a multiplication function as an example: $\X{y} = Mul(\X{x}, \X{p}) \Defs \X{x} \times \X{p}$ where $\X{x}$ is a suspicious input and $\X{p}$ a certain parameter, and $\X{x}, \X{p}, \X{y} \in \mathbb{R}$. Suppose the given output failure mode is $\XF{y}=\VF{h}$ and we are only studying faulty inputs. Although $\X{p}$ is certain and cannot be faulty, its sign still affects the transition of fault from $\X{x}$ to $\X{y}$:
\begin{align}
&(\XF{y}=\VF{h} \wedge \VR{p}>0) \Imply (\XF{x}=\VF{h}) \label{Eq_Multi_11} \\ 
&(\XF{y}=\VF{h} \wedge \VR{p}<0) \Imply (\XF{x}=\VF{l})  \label{Eq_Multi_12}\\
&(\VR{p}=0) \Imply (\XF{y} \neq \VF{h} \wedge \XF{y} \neq \VF{l}) \label{Eq_Multi_13}
\end{align}

As these statements indicate, what we can conclude about $\XF{x}$ based on $\XF{y}$ varies depending on the value of $\X{p}$. Now, consider a scenario in which our prior knowledge about $\X{p}$ is limited to the fact that $\X{p}$ is certain, but we do not know the actual value or sign of this parameter. The only conclusion that we can make will be the following: $\XF{y}=\VF{h} \Imply \XF{x} = \VF{a}$, which obviously provides a different level of information about $\XF{x}$ compared to what we had in \Eq{Eq_Multi_11}--\Eq{Eq_Multi_13}.

FMR uses the premises given about output failure to draw conclusion about input faults. One of the following outcomes may be achieved as a result:
\begin{enumerate}[label=(\roman*)]
\item minimum conditions, expressing the least we can state about the fault at an input, considering what we know about the output fault and the context.
\item certain causes, expressing the conjunctions of input failure modes that, if they occur, will certainly lead to the given output failure mode, regardless of the conditions of the other inputs.
\end{enumerate}

\begin{definition}\label{Def_Certain_Cause}
Let $\X{y}=f(\X{x}, \X{z})$ be a system (or component) function and $\VF{u}, \VF{v} \in \{\VF{h}, \VF{l}, \VF{t}, \VF{f}\}$ two failure modes. We say $\XF{x}=\VF{v}$ is a \emph{certain cause} for $\XF{y}=\VF{u}$ in $f$ if the occurrence of $\XF{x}=\VF{v}$ while $\X{z}$ remains unchanged leads to the occurrence of $\XF{y}=\VF{u}$ regardless of the value of $\X{z}$. That is: $(\XF{x}=\VF{v} \wedge \XF{z}=\VF{m}) \Imply \XF{y}=\VF{u}$.
\end{definition}

\Def{Def_Certain_Cause} requires that the inputs and output of the function are all in an intended state first. This is based on \Assump{Assump_2}. It also considers that part of the inputs remain unchanged during the transition to a faulty state. Thus, the observed change at the output can only be due to the input fault. 

In a general form, let $\X{y}=f(\X{x}_1, .., \X{x}_n)$ be the function under study and $\phi(\XF{x}_1, .., \XF{x}_m)$ an expression on $\XF{x}_1, .., \XF{x}_m$, with $m \leq n$. If $\phi$ is a certain cause for $\XF{y}=\VF{u}$, then:
\begin{align}
(\phi(\XF{x}_1, .., \XF{x}_m) \wedge_{i=m+1}^{n} (\XF{x}_i=\VF{m})) \Imply \XF{y}=\VF{u}
\end{align}

Where $\phi$ is the disjunction of all the certain causes, we can also say:
\begin{align}
((\XF{y}=\VF{u}) \wedge_{i=m+1}^{n} (\XF{x}_i=\VF{m})) \Imply \phi(\XF{x}_1, .., \XF{x}_m)
\end{align}

Certain causes provide a strong basis for reasoning; however, depending on the function, they may not be easy to calculate. In such cases, we may need to use minimum conditions to reason about failures, even though the outcome of such reasoning may not be as strong.

\begin{definition}\label{Def_Cond_Cause}
Let $\X{y}=f(\X{x}, \X{z})$ be a system function and $\VF{u}, \VF{v} \in \{\VF{h}, \VF{l}, \VF{t}, \VF{f}\}$ two failure modes. We say $\XF{x}=\VF{v}$ is a \emph{conditional cause} for $\XF{y}=\VF{u}$ in $f$ if the occurrence of $\XF{x}=\VF{v}$ while $\X{z}$ remains unchanged can lead to $\XF{y}=\VF{u}$ for some values of $\X{z}$.
\end{definition}

Compared to \Def{Def_Certain_Cause} of certain causes, \Def{Def_Cond_Cause} leads to the minimal conditions that \emph{can} result in the nominated output failure. 

We now demonstrate how failure models are derived from function models. The $Or$ function in \Fig{Fig_System1} is defined as: $\X{y} \Defs Or(\X{z}_1, \X{z}_2) \Defs \X{z}_1 \vee \X{z}_2$, with $\X{z}_1, \X{z}_2, \X{y} \in \mathbb{B}$. It follows from this definition that:
\begin{align}
\XF{y}=\VF{f} &\Imply (\VR{y}=\False \wedge \VA{y}=\True) \nonumber \\
&\Imply ((\VR{z}_1=\False \wedge \VR{z}_2=\False) \wedge (\VA{z}_1=\True \vee \VA{z}_2=\True)) \nonumber \\
&\Imply ((\XF{z}_1=\VF{f} \wedge \VR{z}_2=\False) \vee (\XF{z}_2=\VF{f} \wedge \VR{z}_1=\False)) \label{Eq_OR_01} \\
\XF{y}=\VF{t} &\Imply (\VR{y}=\True \wedge \VA{y}=\False) \nonumber \\
&\Imply ((\VR{z}_1=\True \vee \VR{z}_2=\True) \wedge (\VA{z}_1=\False \wedge \VA{z}_2=\False)) \nonumber \\
&\Imply ((\XF{z}_1=\VF{t} \wedge \VA{z}_2=\False) \vee (\XF{z}_2=\VF{t} \wedge \VA{z}_1=\False)) \label{Eq_OR_02} 
\end{align}

The minimum conditions (see \Def{Def_Cond_Cause}) that we can draw from \Eq{Eq_OR_01} and \Eq{Eq_OR_02} are both in disjunctive terms:
\begin{align}
\XF{y}=\VF{f} &\Imply (\XF{z}_1=\VF{f} \vee \XF{z}_2=\VF{f}) \label{Eq_OR_03} \\
\XF{y}=\VF{t} &\Imply (\XF{z}_1=\VF{t} \vee \XF{z}_2=\VF{t}) \label{Eq_OR_04}
\end{align}
This is obviously not a surprise; if the output of $Or$ is faulty, it can only be due to a fault at either input. The certain causes (see \Def{Def_Certain_Cause}), also drawn from \Eq{Eq_OR_01} and \Eq{Eq_OR_02}, provide more information:  
\begin{align}
\XF{y}=\VF{f} &\Imply (\XF{z}_1=\VF{f} \wedge \XF{z}_2=\VF{f}) \label{Eq_OR_05} \\
\XF{y}=\VF{t} &\Imply (\XF{z}_1=\VF{t} \vee \XF{z}_2=\VF{t}) \label{Eq_OR_06}
\end{align}

Statements \Eq{Eq_OR_03} - \Eq{Eq_OR_06} are suitable for value-independent analyses. For value-dependent scenarios, \Eq{Eq_OR_01} and \Eq{Eq_OR_02} should be directly used.

$Gcom$ is another function used in our worked example in \Fig{Fig_System1}. This function is defined as follows: $\X{z}_2 = Gcom(\X{x}, \X{p}_2) \Defs (\X{x} > \X{p}_2)$. From this definition, it easily follows that:
\begin{align}
\XF{z}_2=\VF{f} &\Imply (\XF{x}=\VF{l} \vee \XF{p}_2=\VF{h}) \label{Eq_Gcom_01} \\
\XF{z}_2=\VF{t} &\Imply (\XF{x}=\VF{h} \vee \XF{p}_2=\VF{l}) \label{Eq_Gcom_02}
\end{align}

Similarly, from the definition of function $Lcom$: $\X{z}_1 = Lcom(\X{x}, \X{p}_1) \Defs (\X{x} < \X{p}_1)$, it follows that:
\begin{align}
\XF{z}_1=\VF{f} &\Imply (\XF{x}=\VF{h} \vee \XF{p}_1=\VF{l}) \label{Eq_Lcom_01} \\
\XF{z}_1=\VF{t} &\Imply (\XF{x}=\VF{l} \vee \XF{p}_1=\VF{h}) \label{Eq_Lcom_02}
\end{align}

Note that unlike \Eq{Eq_OR_01} and \Eq{Eq_OR_02} for the $Or$ function, the derived failure models of some functions, such as \Eq{Eq_Gcom_01} - \Eq{Eq_Lcom_02} for $Gcom$ and $Lcom$, may only contain value-independent terms.

\subsection{Composition}\label{Sec_Composition}

Where multiple individual functions are combined to compose a complex system architecture, the global failure behavior of the system can be determined based on the local models that represent the failure behavior of individual functions. What logically relates the local failure models to each other is the interconnections between the component functions. That is, we do not need to define the functional behavior of the whole system to determine its failure behavior. Instead, we can combine the local failure models of individual components, and draw the overall failure model by using rules of logic.

Logically, each component model will be a set of implication statements. The logical statement related to interconnected components can be merged by applying logic rules. In addition, we use the following FMR-specific rule to trim out the always-$True$ and always-$False$ sentences.

\begin{axiom}
A system variable can only be in one state of failure at once. For $\X{x} \in \mathbb{R}$ and $\X{y} \in \mathbb{B}$: 
\begin{align}
&(\XF{x}=\VF{l} \wedge \XF{x}=\VF{h}) \equiv (\XF{x}=\VF{l} \wedge \XF{x}=\VF{m})  \equiv (\XF{x}=\VF{h} \wedge \XF{x}=\VF{m}) \equiv \False  \label{Eq_FMRRule1}\\
&(\XF{y}=\VF{t} \wedge \XF{y}=\VF{f}) \equiv (\XF{y}=\VF{t} \wedge \XF{y}=\VF{m})  \equiv (\XF{y}=\VF{f} \wedge \XF{y}=\VF{m}) \equiv \False  \label{Eq_FMRRule2}\\
&(\XF{x} = \VF{a}) \equiv (\XF{y} = \VF{a}) \equiv \True  \label{Eq_FMRRule3}
\end{align}
\end{axiom}

We now finalize the failure modeling for our worked example in \Fig{Fig_System1}. Suppose we are interested in finding the causes of a $\XF{y}=\VF{t}$ failure and we know that input $\X{x}$ is a `certain' variable. We know from \Eq{Eq_OR_06}, \Eq{Eq_Gcom_02} and \Eq{Eq_Lcom_02} that:
\begin{align}
&\XF{y}=\VF{t} \Imply (\XF{z}_1=\VF{t} \lor \XF{z}_2=\VF{t}) \label{Eq_Ex_Or1} \\
&\XF{z}_2=\VF{t} \Imply \XF{p}_2=\VF{l} \label{Eq_Ex_Gcom1} \\
&\XF{z}_1=\VF{t} \Imply \XF{p}_1=\VF{h} \label{Eq_Ex_Lcom1}
\end{align}

By substituting \Eq{Eq_Ex_Gcom1} and \Eq{Eq_Ex_Lcom1} in \Eq{Eq_Ex_Or1}, we compose the overall model of the system failure: 
\begin{equation}\label{Eq_InFMR}
\XF{y}=\VF{t} \Imply (\XF{p}_1=\VF{h} \lor \XF{p}_2=\VF{l})
\end{equation}

\section{Modeling rules}\label{Sec_CompFunctions}

A component is the smallest part for which we can define a function. This clearly depends on the context of analysis and the suitability of the level of details. For an ordinary car owner, the whole engine will be one component, but a mechanic breaks that component into smaller parts, of which the mechanic can (and needs to) define the functions. Component functions can be distinguished with their following two main features:
\begin{enumerate}[label=(\roman*)]
\item A component function is unbreakable to smaller parts, and is always used as a whole, even if some parts of it are not utilized.
\item The functional behavior of a component is fixed and well defined.
\end{enumerate}

The failure behavior of a component is linked to its functional behavior. We should remember, however, that functional and failure behaviors are expressed in two different spaces. 

We earlier showed FMR failure models for $Not$, $Or$, $Gcom$, and $Lcom$. The detail derivation of these and some more functions is given in the Appendix. 
  In what follows, we set out some common rules that can be used for modeling individual components and we formulate FMR models for common system structures of DNF, CNF, and KooN.

\subsection{Sign, and direction of fault}\label{Sec_Sign_Inv}

Let us begin with the relation between sign of numerical variables and change of direction in failure mode variables. The sign function is defined as follows:
\begin{align} \label{Eq_SignReal}
&sgn(\X{x})=
\begin{cases}
  	\X{x}/|\X{x}| & \text{if $\X{x} \neq 0$}\\
  	0 & \text{if $\X{x} = 0$}
\end{cases} \qquad \text{for $\X{x} \in \mathbb{R}$}
\end{align}

With relation to failure mode variables, we use $\{\NF, null\}$ to express possible changes of direction for faults, with $null$ indicating the same direction and $\NF$ indicating the opposite direction. Thus, if $\XF{x}=\VF{t}$, then $\NF \XF{x}=\VF{f}$; if $\XF{x}=\VF{f}$, $\NF\XF{x}=\VF{t}$; if $\XF{x}=\VF{h}$, $\NF\XF{x}=\VF{l}$; if $\XF{x}=\VF{l}$, $\NF\XF{x}=\VF{h}$; and if $\XF{x}=\VF{m}$, $\NF\XF{x}=\VF{m}$. 

Now, we define the failure mode operator $\Sign{\X{x}}$ to relate $\NF$ to $sgn(\X{x})$:
\begin{align} \label{Eq_Sign}
&\Sign{\X{x}}\equiv
\begin{cases}
  	\NF & \text{if $sgn(\X{x}) = -1$}\\
  	null & \text{otherwise}
\end{cases} \qquad \text{for $\X{x} \in \mathbb{R}$}
\end{align}

\begin{axiom} \label{Axiom_Sign1}
Two consecutive changes of direction means no change: $\NF \NF \equiv null$.
\end{axiom} 

\subsection{Monotonic functions}

The reasoning process in FMR is based on abstracting the values into directions of change. One particularly interesting group of functions that can be well suited for such abstractions is the category of monotonic functions.  

\begin{axiom} \label{Axiom_MonoFun1}
Let $\X{y}\Defs f(\X{x})$ represent a monotonic function. If $f$ is increasing (i.e., $\X{x}' > \X{x}'' \Imply f(\X{x}') \geq f(\X{x}'')$), then $(\XF{y}=\VF{h} \vee \XF{y}=\VF{l}) \Imply \XF{x}=\XF{y}$. If $f$ is decreasing (i.e., $\X{x}'>\X{x}'' \Imply f(\X{x}') \leq f(\X{x}'')$), then $(\XF{y}=\VF{h} \vee \XF{y}=\VF{l}) \Imply \XF{x}=\NF\XF{y}$.\footnote{Arbitrary values $\X{x}'$ and $\X{x}''$ are not to be confused with the `intended value' symbol $\VA{x}$.}
\end{axiom} 

When the inequality $f(\X{x}') \geq f(\X{x}'')$ (or $f(\X{x}') \leq f(\X{x}'')$) is replaced with $f(\X{x}') > f(\X{x}'')$ (or $f(\X{x}') < f(\X{x}'')$), the function is `strictly' monotonic. 

\begin{corollary} \label{Corollary_Mon1}
Let $f: \mathbb{R} \rightarrow \mathbb{R}$ be a continuous strictly monotonic function. Let $\X{x}, \X{y} \in \mathbb{R}$ and $\X{y}\Defs f(\X{x})$. Then: $(\XF{y}=\VF{h} \vee \XF{y}=\VF{l}) \Imply \XF{x}=\Sign{\frac{\partial f}{\partial \X{x}}}\XF{y}$.
\end{corollary} 

Here, $\frac{\partial f}{\partial \X{x}}$ denotes the gradient of $f$. Depending on the sign of this gradient, $\Sign{\frac{\partial f}{\partial \X{x}}}$ determines if $\XF{x}$ and $\XF{y}$ are in the same or opposite direction. 

One particular property of strictly monotonic functions is that the composition of two strictly monotonic functions is itself a strictly monotonic function.

\begin{corollary} \label{Corollary_Mon2}
Let $f, g: \mathbb{R} \rightarrow \mathbb{R}$ be two continuous strictly monotonic functions. Let $\X{x}, \X{y} \in \mathbb{R}$ and $\X{y}\Defs f(g(\X{x}))$. Then: $(\XF{y}=\VF{h} \vee \XF{y}=\VF{l}) \Imply \XF{x}=\Sign{\frac{\partial f}{\partial \X{g}}}\Sign{\frac{\partial g}{\partial \X{x}}}\XF{y}$.
\end{corollary} 

We use these properties of monotonic functions in identifying certain causes.

\begin{lemma} \label{Lemma_Mon1}
Let $\X{y}=f(\X{x}_1, .., \X{x}_n)$ be a function that is strictly monotonic with respect to $\X{x}_i$ (i.e., $\X{x}_i' > \X{x}_i'' \Imply f(\X{x}_1, .., \X{x}_i', .., \X{x}_n) > f(\X{x}_1, .., \X{x}_i'', .., \X{x}_n)$, or $\X{x}_i' > \X{x}_i'' \Imply f(\X{x}_1, .., \X{x}_i', .., \X{x}_n) < f(\X{x}_1, .., \X{x}_i'', .., \X{x}_n)$). Then $\XF{x}_i$ is a certain cause for $\XF{y}$.
\end{lemma}

\begin{proof} 
Follows from the definition of certain causes (see \Def{Def_Certain_Cause}).
\end{proof} 

\begin{corollary} \label{Corollary_Mon3}
Let $\X{y}=f(\X{x}_1, .., \X{x}_n)$ be a continuous strictly monotonic function with respect to each of its individual arguments. Then: $(\XF{y}=\VF{h} \vee \XF{y}=\VF{l}) \Imply \bigvee_{i=1}^{n}(\XF{x}_i=\Sign{\frac{\partial f}{\partial \X{x}_i}}\XF{y})$.
\end{corollary} 

\subsection{Common system structures}\label{Sec_Sel_Funct}

DNF and CNF are two standard forms in which composition rules can be generalized. These normal forms also play an important role in Boolean functions; every Boolean function can be transformed into a DNF or CNF \cite{Ref_226}. Therefore, formulating FMR models for these two normal forms can be helpful in many other scenarios in which Boolean functions are the subject of the study. 

\paragraph{DNF:} 
Let $\Bold{\X{x}}=[\X{x}_{lk}]_{L\times K}$ be a matrix of Boolean variables. We define the DNF function $\X{y} \Defs DNF(\Bold{\X{x}})$ as follows:
\begin{align}
&\X{y} \Defs DNF(\Bold{\X{x}}) \Defs \bigvee_{l=1}^{L} \X{y}_l \Defs \bigvee_{l=1}^{L} \bigwedge_{k=1}^{K} \X{x}_{lk}, \textit{ with }~ \X{y}_l \Defs \bigwedge_{k=1}^{K} \X{x}_{lk}
\end{align}

Based on \Def{Def_Certain_Cause}, the certain causes of fault at $\X{y}$ will be as follows:
\begin{eqnarray}
\{ \bigvee_{l} \bigwedge_{k} \XF{x}_{lk}=\XF{y}\} &~~DNF~~& \{\XF{y}=\VF{t} \} \label{Eq_DNF_gh6} \\
\{ \bigwedge_{l} \bigvee_{k} \XF{x}_{lk}=\XF{y}\} &~~DNF~~& \{\XF{y}=\VF{f} \} \label{Eq_DNF_gh7}
\end{eqnarray}
And the minimum conditions will be as follows:
\begin{eqnarray}
\{ \bigvee_{l, k} \XF{x}_{lk}=\XF{y}\} &~~DNF~~& \{\XF{y}=\VF{t} \vee \XF{y}=\VF{f}\} \label{Eq_DNF_CD0}
\end{eqnarray}

\textit{Proof:} To prove the minimum conditions, let us begin with the disjunction $\X{y}=\lor_l y_l$. Recall that the reported and intended values at $\X{y}$ are denoted by $\VR{y}$ and $\VA{y}$, and the reported and intended values at $\X{y}_l$ by $\VR{y}_l$ and $\VA{y}_l$. From $\XF{y}=\VF{f}$, it follows that:
\begin{align}
\XF{y}=\VF{f} &\Imply (\VR{y}=\False \wedge \VA{y}=\True) \Imply \bigwedge_{j=1}^{j=L} (\VR{y}_j=\False) \wedge \bigvee_{l=1}^{l=L} (\VA{y}_l=\True) \nonumber \\
&\Imply \bigvee_{l=1}^{l=L} (\XF{y}_l=\VF{f}) \bigwedge_{\substack{j=1 \\ j \neq l}}^{j=L} (\VR{y}_j=\False) \Imply \bigvee_{l=1}^{l=L} (\XF{y}_l=\VF{f}) \label{Eq_DNF_D1} \\
\XF{y}=\VF{t} &\Imply (\VR{y}=\True \wedge \VA{y}=\False) \Imply \bigvee_{l=1}^{l=L} (\VR{y}_l=\True) \wedge \bigwedge_{j=1}^{j=L} (\VA{y}_l=\False) \nonumber \\
&\Imply \bigvee_{l=1}^{l=L} (\XF{y}_l=\VF{t}) \bigwedge_{\substack{j=1 \\ j \neq l}}^{j=L} (\VA{y}_j=\False) \Imply \bigvee_{l=1}^{l=L} (\XF{y}_l=\VF{t}) \label{Eq_DNF_D2}
\end{align}
Further, each conjunctive clause in DNF is defined by $\X{y}_l=\wedge_k \X{y}_{lk}$. Similar to the disjunctive part, we can say the following:
\begin{align}
\XF{y}_l=\VF{f} &\Imply (\VR{y}_l=\False \wedge \VA{y}_l=\True) \Imply \bigvee_{k=1}^{k=K} (\VR{x}_{lk}=\False) \wedge \bigwedge_{j=1}^{j=K} (\VA{x}_{lj}=\True) \nonumber \\
&\Imply \bigvee_{k=1}^{k=K} (\XF{x}_{lk}=\VF{f}) \bigwedge_{\substack{j=1 \\ j \neq k}}^{j=K} (\VA{x}_{lj}=\True) \Imply \bigvee_{k=1}^{k=K} (\XF{x}_{lk}=\VF{f}) \label{Eq_DNF_C1} \\
\XF{y}_l=\VF{t} &\Imply (\VR{y}_l=\True \wedge \VA{y}_l=\False) \Imply \bigwedge_{j=1}^{j=K} (\VR{x}_{lj}=\True) \wedge \bigvee_{k=1}^{k=K} (\VA{x}_{lk}=\False) \nonumber \\
&\Imply \bigvee_{k=1}^{k=K} (\XF{x}_{lk}=\VF{t}) \bigwedge_{\substack{j=1 \\ j \neq k}}^{j=K} (\VR{x}_{lj}=\True) \Imply \bigvee_{k=1}^{k=K} (\XF{x}_{lk}=\VF{t}) \label{Eq_DNF_C2} 
\end{align}

By substituting \Eq{Eq_DNF_C1} in \Eq{Eq_DNF_D1}, and \Eq{Eq_DNF_C2} in \Eq{Eq_DNF_D2}, we will have the following:
\begin{align}
\XF{y}=\VF{f} &\Imply \bigvee_{l=1}^{l=L} \bigvee_{k=1}^{k=K} (\XF{x}_{lk}=\VF{f}) \label{Eq_DNF_C3} \\
\XF{y}=\VF{t} &\Imply \bigvee_{l=1}^{l=L} \bigvee_{k=1}^{k=K} (\XF{x}_{lk}=\VF{t}) \label{Eq_DNF_C4} 
\end{align}
which together prove \Eq{Eq_DNF_CD0}. 

We now prove the certain causes by using composition rules. Let us begin with the disjunction $\X{y}=\lor_l y_l$:
\begin{align}
\XF{y}=\VF{f} &\Imply (\VR{y}=\False \wedge \VA{y}=\True)  \nonumber \\
&\Imply \bigwedge_{l=1}^{l=L} (\VR{y}_l=\False) \wedge \bigvee_{j=1}^{j=L} (\VA{y}_j=\True) \nonumber \\
&\Imply \bigvee_{j=1}^{j=L} (\VA{y}_j=\True) \bigwedge_{l=1}^{l=L} (\VR{y}_l=\False \wedge (\VA{y}_l=\False \vee \VA{y}_l=\True)) \nonumber \\
&\Imply \bigvee_{j=1}^{j=L} (\VA{y}_j=\True) \bigwedge_{l=1}^{l=L} ((\VR{y}_l=\False \wedge \VA{y}_l=\False) \vee (\VR{y}_l=\False \wedge \VA{y}_l=\True)) \label{Eq_DNF_gh1}
\end{align}

To simplify the formulation, let $A_l \Defs (\VR{y}_l=\False \wedge \VA{y}_l=\False)$, $B_l \Defs (\VR{y}_l=\False \wedge \VA{y}_l=\True)$ and $C \Defs \bigvee_{j=1}^{j=L} (\VA{y}_j=\True)$. Thus, the conclusion part of \Eq{Eq_DNF_gh1} will be the following:
\begin{align}
&C \bigwedge_{l=1}^{l=L} (A_l \vee B_l) \equiv (C \bigwedge_{l=1}^{l=L} A_l) \vee .. \vee (C \bigwedge_{m \neq n} A_m B_n) \vee .. \vee (C \bigwedge_{l=1}^{l=L} B_l) \label{Eq_DNF_gh2}
\end{align}

The certain causes of a DNF, if any, should come from the conjunction of failure modes included in \Eq{Eq_DNF_gh2}. So, let us examine each constituting term separately. First, the term $C \bigwedge_{l=1}^{l=L} A_l$ does not hold; because it requires $\bigwedge_{l=1}^{l=L} (\VA{y}_l=\False)$, which contradicts the premise $\bigvee_{l=1}^{l=L} (\VA{y}_l=\True)$. 

The middle terms in \Eq{Eq_DNF_gh2} are combinations of faulty and non-faulty variables. The non-faulty variables (in $A_m$s) remain unchanged during the time when the failure occurs. Thus, we want to see if the conjunction of faulty variables (in $\bigwedge_n B_n$) is consistent with \Def{Def_Certain_Cause} because it is only in that case that $\bigwedge_n B_n$ can be a certain cause. We know that for $\bigwedge_{n} B_n$ to be a certain cause, $\bigwedge_{n} B_n$ needs to cause a change of output state independently of the status of other $\X{y}_l$ variables. That is, regardless of the reported values at $\X{y}_l$s, the occurrence of $\bigwedge_{n} B_n$ should always lead to the same change of state at the DNF output. This includes the case where there is a clause with output state $\VA{y}_l=\True \wedge \VR{y}_l=\True$, which is, obviously, not possible because if $\X{y}_l=\True$ before, during, and after the occurrence of failure, then the output $\X{y}$ will remain $\True$ throughout, and no change will be observed at the output.\footnote{This is similar to our discussion in \Sec{Sec_Causation}, where we used truth table to prove a small AND gate with two inputs.} 

The last term $C \bigwedge_{l=1}^{l=L} B_l$, however, depicts a certain cause: $\bigwedge_{l=1}^{l=L} (\XF{y}_l=\VF{f})$. If all the $\X{y}_l$ variables fail to $\XF{y}_l=\VF{f}$, the output of the disjunction will certainly fail to $\XF{y}=\VF{f}$. Thus, the \Eq{Eq_DNF_gh1} can be simplified as follows:
\begin{align}
&C \bigwedge_{l=1}^{l=L} B_l \equiv \bigvee_{j=1}^{j=L} (\VA{y}_j=\True) \bigwedge_{l=1}^{l=L} (\VR{y}_l=\False \wedge \VA{y}_l=\True) \equiv \bigwedge_{l=1}^{l=L} (\XF{y}_l=\VF{f}) \label{Eq_DNF_gh3}
\end{align}

We now use the conjunctive part of DNF: $\X{y}_l = \bigwedge_{k=1}^{K} \X{x}_{lk}$, to identify the certain causes of $\XF{y}_l=\VF{f}$. The result can then be substituted in \Eq{Eq_DNF_gh3}. From $\XF{y}_l=\VF{f}$, we have:
\begin{align}
\XF{y}_l=\VF{f} &\Imply (\VR{y}_l=\False \wedge \VA{y}_l=\True)  \nonumber \\
&\Imply \bigvee_{k=1}^{k=K} (\VR{x}_{lk}=\False) \wedge \bigwedge_{j=1}^{j=K} (\VA{x}_{lj}=\True) \nonumber \\
&\Imply \bigvee_{k=1}^{k=K} (\XF{x}_{lk}=\VF{f}) \bigwedge_{\substack{j=1 \\ j \neq k}}^{j=K} (\VA{x}_{lj}=\True) \nonumber \\
&\Imply \bigvee_{k=1}^{k=K} (\XF{x}_{lk}=\VF{f}) \label{Eq_DNF_gh4}
\end{align}

Substituting \Eq{Eq_DNF_gh4} in \Eq{Eq_DNF_gh3} proves one part of the proposed model (i.e., failure scenario \Eq{Eq_DNF_gh7}). The other part of the proposed model (i.e., failure scenario \Eq{Eq_DNF_gh6}) can be proved in a similar manner. We begin with the disjunctive part of the DNF:
\begin{align}
\XF{y}=\VF{t} &\Imply (\VR{y}=\True \wedge \VA{y}=\False)  \nonumber \\
&\Imply \bigvee_{l=1}^{l=L} (\VR{y}_{l}=\True) \wedge \bigwedge_{j=1}^{j=L} (\VA{y}_{j}=\False) \nonumber \\
&\Imply \bigvee_{l=1}^{l=L} (\XF{y}_{l}=\VF{t}) \bigwedge_{\substack{j=1 \\ j \neq k}}^{j=K} (\VA{y}_{j}=\False) \nonumber \\
&\Imply \bigvee_{l=1}^{l=L} (\XF{y}_{l}=\VF{t}) \label{Eq_DNF_ghf4}
\end{align}

We now use the conjunctive part of DNF: $\X{y}_l = \bigwedge_{k=1}^{K} \X{x}_{lk}$, to identify the certain causes of $\XF{y}_l=\VF{t}$, which we will later substitute in \Eq{Eq_DNF_ghf4}:
\begin{align}
\XF{y}_l=\VF{t} &\Imply (\VR{y}_l=\True \wedge \VA{y}_l=\False)  \nonumber \\
&\Imply \bigwedge_{k=1}^{k=K} (\VR{x}_{lk}=\True) \wedge \bigvee_{j=1}^{j=K} (\VA{x}_{lj}=\False) \nonumber \\
&\Imply \bigvee_{j=1}^{j=K} (\VA{x}_{lj}=\False) \bigwedge_{k=1}^{k=K} (\VR{x}_{lk}=\True \wedge (\VA{x}_{lk}=\True \vee \VA{x}_{lk}=\False)) \nonumber \\
&\Imply \bigvee_{j=1}^{j=K} (\VA{x}_{lj}=\False) \bigwedge_{k=1}^{k=K} ((\VR{x}_{lk}=\True \wedge \VA{x}_{lk}=\True) \vee (\VR{x}_{lk}=\True \wedge \VA{x}_{lk}=\False)) \label{Eq_DNF_ghf5}
\end{align}

Let $A_{lk} \Defs (\VR{x}_{lk}=\True \wedge \VA{x}_{lk}=\True)$, $B_{lk} \Defs (\VR{x}_{lk}=\True \wedge \VA{x}_{lk}=\False)$ and $C \Defs \bigvee_{j=1}^{j=K} (\VA{x}_{lj}=\False)$. Thus, the conclusion part of \Eq{Eq_DNF_ghf5} will be the following:
\begin{align}
&C \bigwedge_{k=1}^{k=K} (A_{lk} \vee B_{lk}) \equiv (C \bigwedge_{k=1}^{k=K} A_{lk}) \vee .. \vee (C \bigwedge_{m \neq n} A_m B_n) \vee .. \vee (C \bigwedge_{k=1}^{k=K} B_{lk}) \label{Eq_DNF_ghf6}
\end{align}
Similar to \Eq{Eq_DNF_gh2}, the only possible certain causes from \Eq{Eq_DNF_ghf6} will be the following:
\begin{align}
&C \bigwedge_{k=1}^{k=K} B_{lk} \equiv \bigvee_{j=1}^{j=K} (\VA{x}_{lj}=\False) \bigwedge_{k=1}^{k=K} (\VR{x}_{lk}=\True \wedge \VA{x}_{lk}=\False) \equiv \bigwedge_{k=1}^{k=K} (\XF{x}_{lk}=\VF{t}) \label{Eq_DNF_ghfx7}
\end{align}
which, with respect to certain causes, means the following:
\begin{align}
\XF{y}_l=\VF{t} &\Imply \bigwedge_{k=1}^{k=K} (\XF{x}_{lk}=\VF{t}) \label{Eq_DNF_ghf7}
\end{align}
Substituting \Eq{Eq_DNF_ghf7} in \Eq{Eq_DNF_ghfx7} proves the failure scenario \Eq{Eq_DNF_gh6}.

\paragraph{CNF:} 
Similar to a DNF, we define the CNF function $\X{y} \Defs CNF(\Bold{\X{x}})$ as follows:
\begin{align}
&\X{y} \Defs CNF(\Bold{\X{x}}) \Defs \bigwedge_{l=1}^{L} \X{y}_l \Defs \bigwedge_{l=1}^{L} \bigvee_{k=1}^{K} \X{x}_{lk}, \textit{ with } \X{y}_l \Defs \bigvee_{k=1}^{K} \X{x}_{lk}
\end{align}
Based on \Def{Def_Certain_Cause}, the certain causes of fault at $\X{y}$ will be as follows:
\begin{eqnarray}
\{ \bigwedge_{l} \bigvee_{k} \XF{x}_{lk}=\XF{y}\} &~~CNF~~& \{\XF{y}=\VF{t} \} \label{Eq_CNF_gh6} \\
\{\bigvee_{l} \bigwedge_{k} \XF{x}_{lk}=\XF{y}\} &~~CNF~~& \{\XF{y}=\VF{f} \} \label{Eq_CNF_gh7}
\end{eqnarray}
And the minimum conditions will be as follows:
\begin{eqnarray}
\{ \bigvee_{l, k} \XF{x}_{lk}=\XF{y}\} &~~CNF~~& \{\XF{y}=\VF{t} \vee \XF{y}=\VF{f}\}\label{Eq_CNF_CD0}
\end{eqnarray}

\textit{Proof:} Similar to the DNF function.

\paragraph{KooN:} 
A $KooN$ system is a composition of $n$ independent and identical components, where $k$ out of $n$ components need to function for the overall system to function. This system can be modeled as follows. For $\X{y}, \X{x}_1, .., \X{x}_n \in \mathbb{B}$, where $\X{x}_1, .., \X{x}_n$ are suspicious variables:
\begin{align}
&\X{y} \Defs  KooN(\X{x}_1, .., \X{x}_n, k) = (\X{x}_1 \wedge .. \wedge \X{x}_{k-1} \wedge \X{x}_k) \vee (\X{x}_1 \wedge ..\nonumber \\
& \wedge \X{x}_{k-1} \wedge \X{x}_{k+1}) \vee .. \vee (\X{x}_{n-k+1} \wedge .. \wedge \X{x}_{n-1} \wedge \X{x}_n) 
\end{align}

In a $KooN$ function, the output will be $\X{y}=\True$ if $k$ inputs are $\True$, and the output will be $\X{y}=\False$ otherwise. The certain causes of this function are as follows:
\begin{align}
\{&(\XF{x}_1=\VF{t} \wedge .. \wedge \XF{x}_{k-1}=\VF{t} \wedge \XF{x}_k=\VF{t}) \vee (\XF{x}_1=\VF{t} \wedge .. \wedge \XF{x}_{k-1}=\VF{t} \wedge \XF{x}_{k+1}=\VF{t}) \nonumber\\& \vee .. \vee (\XF{x}_{n-k+1}=\VF{t} \wedge .. \wedge \XF{x}_{n-1}=\VF{t} \wedge \XF{x}_n=\VF{t})\} ~~~~KooN~~~~ \{\XF{y}=\VF{t} \} \label{Eq_KooN_01}\\
\{&(\XF{x}_1=\VF{f} \wedge .. \wedge \XF{x}_{n-k}=\VF{f} \wedge \XF{x}_{n-k+1}=\VF{f}) \vee (\XF{x}_1=\VF{t} \wedge .. \wedge  \XF{x}_{n-k}=\VF{f} \wedge \XF{x}_{n-k+2}=\VF{f}) \nonumber \\& \vee .. \vee  (\XF{x}_k=\VF{f} \wedge .. \wedge \XF{x}_{n-1}=\VF{f} \wedge \XF{x}_n=\VF{f})\} ~~~~KooN~~~~ \{\XF{y}=\VF{f} \} \label{Eq_KooN_02}
\end{align}

\textit{Proof:} Follows from the proof of DNF. Alternatively, truth tables can be used for specific values of $k$ and $n$ (see Appendix for truth-table-based proof for the $And$ function).

~

Our discussion around the DNF, CNF, and KooN functions highlights a few interesting facts:

\begin{enumerate}[label=(\roman*)]
\item First, if the output of a DNF or CNF is deviated in $\VF{f}$ or $\VF{t}$ direction, at least one of its literals is deviated in the same direction (see \Eq{Eq_DNF_CD0} and \Eq{Eq_CNF_CD0}).
\item More generally, if the output of a Boolean function is deviated in $\VF{f}$ or $\VF{t}$ direction, at least one of its inputs is deviated in the same direction; because every Boolean function can be converted to a DNF or CNF.
\item We also know that the certain causes in DNF and CNF systems follow the architecture of the system. That is, the certain causes for an output failure $\XF{y}=\VF{t}$ in a DNF (or CNF) function are expressed by the same DNF (or CNF) used for its success $\X{y}=\True$, if we just replace the literals $\X{x}_{lk}$ with failure modes $\XF{x}_{lk}=\VF{t}$. This is based on \Eq{Eq_DNF_gh6}, \Eq{Eq_DNF_gh7}, \Eq{Eq_CNF_gh6}, and \Eq{Eq_CNF_gh7}. Conversely, for the output failure $\XF{y}=\VF{f}$, the dual architecture CNF (or DNF) should be used for failure modeling, while the literals $\X{x}_{lk}$ are replaced with failure modes $\XF{x}_{lk}=\VF{f}$.
\item  In a system that needs at least $k$ out of $n$ of its components for functioning, the system can fail in two ways: commission of $k$ components resulting in a commission at the system output, and omission of $n-k+1$ components resulting in an omission at the system output (see \Eq{Eq_KooN_01} and \Eq{Eq_KooN_02}).
\end{enumerate}

\subsection{Duality and equality}\label{Sec_DualEqual}

Where the functional behaviors of two systems are the same, one may also expect to observe the same failure behaviors from the two systems. Conversely, where two system functions `negate' each other, one may expect to see opposite failure behaviors from the two systems. In this section, we examine these questions. First, we address the question of equality.

\begin{theorem}\label{Thm_Equality}
Given two functions $\X{y}=f(\X{x}_1, .., \X{x}_n)$ and $\X{y}=g(\X{x}_1, .., \X{x}_n)$ and failure mode $\VF{u}\in \{\VF{f}, \VF{t}, \VF{l}, \VF{h}\}$, if $\XF{y}=\VF{u}$ is reachable in $f$ and $g$, and if $f$ and $g$ are equal (i.e., if for any given input $\X{x}_1, .., \X{x}_n$ we have $f(\X{x}_1, .., \X{x}_n)=g(\X{x}_1, .., \X{x}_n)$), then the certain causes of $f$ and $g$ for $\XF{y}=\VF{u}$ are identical. 
\end{theorem}

\begin{proof} 
As per \Def{Def_Certain_Cause}, a certain cause is a conjunction of failure modes of a number of inputs while the other inputs are unchanged. Let $\phi (\X{x}_1, .., \X{x}_m)$ represent a certain cause for $\XF{y}=\VF{u}$ in $f$. Thus:
\begin{align}
\phi \wedge_{i>m} (\XF{x}_i =\VF{m}) &\Imply \XF{y}=\VF{u} \textit{, due to f}\nonumber \\
&\Imply \MD(f(\VR{x}_1, .., \VR{x}_n), f(\VA{x}_1, .., \VA{x}_n))=\VF{u} \nonumber \\
&\Imply \MD(g(\VR{x}_1, .., \VR{x}_n), g(\VA{x}_1, .., \VA{x}_n))=\VF{u} \nonumber \\
&\Imply \XF{y}=\VF{u} \textit{, due to g}
\end{align}
which means $\phi (\X{x}_1, .., \X{x}_m)$ is also a certain cause for $\XF{y}=\VF{u}$ in $g$.
\end{proof} 

We now discuss the `duality' in failure models. In standard logic, the dual of a Boolean function $f$ is depicted by $f^d$, which is composed by exchanging all the $\wedge$ and $\vee$ operators and negating all the literals in $f$ \cite{Ref_226}. We first introduce a similar notation for failure models. 
 
\begin{definition}
Given is $\phi(\XF{x}_1, .., \XF{x}_n)$, a Boolean expression on failure mode variables $\XF{x}_j$s. The \emph{failure mode dual} of $\phi$ is depicted by $\phi^D(\XF{x}_1, .., \XF{x}_n)$, in which all the $\wedge$ and $\vee$ operators are exchanged and all the failure mode values inverted by applying the $\NF$ operator introduced in \Sec{Sec_Sign_Inv}.  
\end{definition}

As an example, for $\phi \equiv (\XF{x}_1=\VF{h} \wedge \XF{x}_2=\VF{l} \wedge (\XF{x}_1=\VF{m} \vee \XF{x}_3=\VF{h}))$, its failure mode dual expression will be $\phi^D \equiv (\XF{x}_1=\VF{l} \vee \XF{x}_2=\VF{h} \vee (\XF{x}_1=\VF{m} \wedge \XF{x}_3=\VF{l}))$.

\begin{theorem}\label{Theorem_Duality}
Given are failure mode $\VF{u}\in \{\VF{f}, \VF{t}\}$ and Boolean function $\X{y}=f(\X{x}_1, .., \X{x}_n)$. If $\phi(\XF{x}_1, .., \XF{x}_m)$ is a certain cause for $\XF{y}=\VF{u}$, then $\phi^D(\XF{x}_1, .., \XF{x}_m)$ is a certain cause for $\XF{y}=\NF\VF{u}$.
\end{theorem}

\begin{proof}
We proved the failure models for DNF and CNF, in which this duality rule is apparent (see \Eq{Eq_DNF_gh6}, \Eq{Eq_DNF_gh7}, \Eq{Eq_CNF_gh6}, and \Eq{Eq_CNF_gh7}). Every Boolean function can be converted to a DNF (or CNF). Let $g$ be the DNF equivalent of $f$. According to \Theorem{Thm_Equality}, the certain causes of $f$ and $g$ are identical. Therefore, we can conclude that the duality rule applies to all Boolean functions.
\end{proof}

\begin{corollary}
Given Boolean function $f$ and its logical dual $f^d$, the certain causes in $f^d$ are the failure mode duals of the certain causes in $f$.
\end{corollary}

\begin{proof} 
Let $\X{y}=f(\X{x}_1, .., \X{x}_n)$ and $\X{y}^d=f^d(\X{x}_1, .., \X{x}_n)$. Let $\phi(\XF{x}_1, .., \XF{x}_m)$ depict a certain cause for $\XF{y}=\VF{t}$ in $f$, with $m\leq n$. Thus:
\begin{align}
\phi(\XF{x}_1, .., \XF{x}_m) \wedge_{i=m+1}^{i=n}(\XF{x}_1=\VF{m}) &\Imply \XF{y}=\VF{t} \nonumber \\
&\Imply (\VR{y}=\True \wedge \VA{y}=\False) \nonumber \\
&\Imply (\VRD{y}=\False \wedge \VAD{y}=\True) \nonumber \\
&\Imply \XFD{y}=\VF{f}
\end{align} 
which means $\phi$ is a certain cause for $\XFD{y}=\VF{f}$. From \Theorem{Theorem_Duality}, it follows that $\phi^D$ is a certain cause for $\XFD{y}=\VF{t}$, and that proves this corollary for output failure mode $\VF{t}$. Similarly, we can show that if $\phi$ is a certain cause for $\XF{y}=\VF{f}$, then $\phi^D$ is a certain cause for $\XFD{y}=\VF{t}$. Thus, the corollary is proven for both $\VF{f}$ and $\VF{t}$.
\end{proof}

\section{Discussion}\label{Sec_Discussion}

\paragraph{Loops:}

Some practical systems use loops in their architectures. A loop connects the output of a function back to its own input through direct or indirect connections. We discuss the effect of loops on the transition of faults.

Consider the looped system defined as follows:
\begin{align}
&\X{y}=f(\Bold{\X{x}}, \X{y}) \label{Eq_Loop_Sys_Fun1}
\end{align}
where $\Bold{\X{x}}=[\X{x}_1 ... \X{x}_n]$ is the system input vector, and the argument $\X{y}$ depicts a loop to the same output $\X{y}$. Considering the two moments in time $t_1$ and $t_2$ such that $t_2=t_1-dt$, the system equation \Eq{Eq_Loop_Sys_Fun1} can be written as follows:
\begin{align}
&\X{y}(t_2)=f(\Bold{\X{x}}(t_1), \X{y}(t_1)) \label{Eq_Loop_Sys_Fun2}
\end{align}

Suppose a failure is observed at $t_2$ (i.e., $\XF{y}(t_2) \neq \VF{m}$ and $\VR{y}(t_2) \neq \VA{y}(t_2)$). Therefore:
\begin{align}
&f(\Bold{\VR{x}}(t_1), \VR{y}(t_1)) \neq f(\Bold{\VA{x}}(t_1), \VA{y}(t_1))
\end{align}

From \Assump{Assump_2}, we know that the system is in an intended state at $t_1$, which means $\VR{y}(t_1) = \VA{y}(t_1)$ and thus $\XF{y}(t_1) = \VF{m}$. This simply means that the only possible cause for $\XF{y}(t_2) \neq \VF{m}$ could be $\Bold{\XF{x}}(t_1)$ and not $\XF{y}(t_1)$. That is, the loop does not affect the output failure.

In practice, the relationship between $\XF{y}(t_2)$ and $\Bold{\XF{x}}(t_1)$ and $\XF{y}(t_1)$ is always expressed in the form of a Boolean expression, comprising literals such as $\XF{x}_j(t_1)=\VF{v}$ and $\XF{y}(t_1)=\VF{u}$. Given that we know for certain that $\XF{y}(t_1)=\VF{m}$, all sentences like $\XF{y}(t_1)=\VF{u}$ in the final expression, where $\VF{u} \neq \VF{m}$, can be replaced with certain $False$, and all $\XF{y}(t_1)=\VF{m}$ sentences can be replaced with certain $True$. These replacements will help further simplify the Boolean expression and reduce it to an expression based on $\Bold{\XF{x}}(t_1)$ only.

The \emph{value} of a loop connection, however, can affect the transition of faults where we choose to take value-dependent reasoning. In \Eq{Eq_Multi_11}-\Eq{Eq_Multi_13}, we showed how the sign of a certain variable could affect the reasoning about the failure mode of another (suspicious) variable. As another example, a correct $\False$ at one input of an $And$ function will block all the other faulty inputs from affecting the output. 

\paragraph{Change and impact:}

In some analyses, we may be interested in the impact of a particular variable on the overall system output failure. Consider a system in which the settings of parameters can be adjusted. When an output failure is observed, one may want to see if altering a parameter can potentially fix the output failure. In this section, we introduce an `impact index' that can be useful in such scenarios. First, we introduce a \emph{comparison} function that can help measure the difference between two given failure modes:

\begin{definition}\label{Def_CompFunction}
Let $\XF{x}$ and $\XF{z}$ be two failure mode variables such that $\XF{x}, \XF{z} \in \{\VF{h}, \VF{m}, \VF{l}\}$ (or $\XF{x}, \XF{z} \in \{\VF{t}, \VF{m}, \VF{f}\}$). The \emph{failure mode comparison function} defined below measures the difference between $\XF{x}$ and $\XF{z}$:
\begin{align}
&cmp(\XF{x}, \XF{z}) \Defs 
\begin{cases}
  	\V{1} \quad~~~if~ \XF{x}=\VF{h} ~and~ \XF{z}=\VF{l}, ~or~ \XF{x}=\VF{t} ~and~ \XF{z}=\VF{f}\\
  	\V{0.5} \quad if~ \XF{z}=\VF{m},~and~ \XF{x}=\VF{h} ~or~ \XF{x}=\VF{t}\\
  	\V{0} \quad~~~ if~ \XF{x}=\XF{z}\\
  	\V{-0.5} \quad if~ \XF{z}=\VF{m}, ~and~ \XF{x}=\VF{l} ~or~ \XF{x}=\VF{f}\\
  	\V{-1} \quad~~if~ \XF{x}=\VF{l} ~and~ \XF{z}=\VF{h}, ~or~ \XF{x}=\VF{f} ~and~ \XF{z}=\VF{t}\\
\end{cases}
\end{align}
\end{definition}

The impact of a variable on the output failure can then be defined as follows.

\begin{definition}\label{Def_Impact}
Let $\X{y}=f(\X{x}_1, .., \X{x}_n)$ be a system function, $\XF{y}$ the failure mode at output $\X{y}$, and $\VR{v}$ and $\VR{w}$ two possible assignments to $\X{x}_j$. The \emph{impact} of a change from $\X{x}_j=\VR{v}$ to $\X{x}_j=\VR{w}$ on the output failure mode $\XF{y}$ is measured by the following:
\begin{align}\label{Eq_Imp}
&I_{\X{x}_j}\Defs |cmp(\XF{y}_v, \XF{y}_w)|, \nonumber \\&~where~ \XF{y}_v = \XF{y}\big|_{\X{x}_j=\VR{v}} ~and~ \XF{y}_w = \XF{y}\big|_{\X{x}_j=\VR{w}}
\end{align}
\end{definition}

The impact index indicates how effectively the change of the current value at $\X{x}_j$ can impact the occurrence of a failure mode at $\X{y}$. 

The measure we introduced in \Eq{Eq_Imp} is useful in systems with single outputs. Where the system produces multiple outputs (e.g., where the system output is $\Bold{\X{y}}=[\X{y}_1 ... \X{y}_m]$), a collective measure can be used to show the impact as follows:
\begin{align}\label{Eq_Imp1}
&I_{\X{x}_j}\Defs \sum_{i=1}^{i=m}|cmp(\XF{y}_{iv}, \XF{y}_{iw})|, \nonumber \\&~where~ \XF{y}_{iv} = \XF{y}_i\big|_{\X{x}_j=\VR{v}} ~and~ \XF{y}_{iw} = \XF{y}_i\big|_{\X{x}_j=\VR{w}}
\end{align}

The impact measure can be particularly useful when the impact of a parameter change is to be evaluated. Where an output failure is an indication of an incorrect parameter and the correct value of the parameter is unknown, one possible way to fix the problem is to propose a new setting for the parameter. In such scenarios, the impact index can help to evaluate the effectiveness of the new setting on output failure.

\paragraph{Application:}\label{Sec_Application}

Our focus in this paper is on the theoretical aspects of FMR. However, we have also realized these concepts in the form of an analysis tool. With respect to SIS programs, the tool has already been used in multiple projects in the power industry. The FMR tool imports the SIS program, analyzes it like a function model of a system, and identifies the possible input and/or parameter faults. We have demonstrated this in detail in the previous works (e.g., in \cite{Ref_225_0,Ref_295_0}). At its current capacity, the tool is capable of analyzing models of cyber physical systems too. We are currently working on a modeling project for a road tunnel safety scenario where the tool is to analyze failures in a complex tunnel ventilation system.

\section{Related works}\label{Sec_RelatedWorks}

FMR was not created to expand or enhance an existing method. It is therefore hard to discuss its relative advantages or limitations in direct comparison with another existing method. However, we can still compare some of the concepts that are commonly utilized by both FMR and the other methods.   

In the previous works on FMR \cite{Ref_200_0,Ref_225_0,Ref_230_0,Ref_295_0}, we compared various aspects of FMR with  FTA \cite{Ref_176,Ref_168}, FMEA \cite{Ref_181}, inference-based diagnostics \cite{Ref_228,Ref_233,Ref_227}, Abstract Interpretation \cite{Ref_235,Ref_297}, Qualitative Reasoning \cite{Ref_232}, component-based methods \cite{Ref_172,Ref_163,Ref_174,Ref_140,Ref_143}, diagnosis methods \cite{Ref_227,Ref_228} and program debugging \cite{Ref_260,Ref_259}.

One particular area with which FMR does \emph{not} have much in common is formal verification. Model Checking \cite{Ref_335}, for instance, is a method for checking whether a model of a system meets a given specification. FMR does not verify a model against a specification, nor does it model a system function. It reads the functional behavior and extracts the failure behavior as it actually is, without making any judgment. It will be then up to the engineers to decide if that identified behavior is acceptable. Similarly, FMR is not to be compared with Model-Based Safety Analysis (MBSA) \cite{Ref_306} because as MBSA does not derive the failure mode from function model.

The closest work to FMR's application in SIS programs is Software FTA \cite{Ref_310}. In simple terms, a software fault tree is a composition of a set of smaller fault trees, each of which represents the failure behavior of a small piece of code or an instruction. The overall fault tree represents the failure behavior of the whole program. The main difference between FMR and SFTA is that SFTA is based on the standard FTA whereas FMR is a method based on abstract deviations at variables (i.e., failure modes). Also, FMR is a method for generic composition of functions whereas SFTA is used solely for failures related to the execution of software programs. 

Despite the differences, both methods address fault propagation, which is also shared with other methods, such as FFIP \cite{Ref_303}. FFIP models the hardware and software components based on their \emph{functional} behavior. An FFIP model is used to analyze the propagation of failure through the (function) model, determining the final failure at the system level. Failure scenarios are then simulated to assess the effect of failure in the design, which is different to what FMR does. 

\section{Conclusion}\label{Sec_Conclusion}

We generalized and extended the FMR formalism for its use in generic systems. Our main objective was to establish the mathematical foundation that is required for failure reasoning, independent from specific application contexts. The framework provided here can be used by safety analysts for all failure mode reasoning scenarios, including SIS programs. Potential future work include implementation of FMR in more industrial cases, inclusion of time delays in FMR models as parameters, and extension of FMR for system architectural errors.  

\bibliographystyle{splncs04}
\bibliography{References}

\clearpage

\appendix

\section{Failure models for selected functions}

In this appendix, we propose FMR models for a selected group of basic functions, and we demonstrate their correctness. We will cover a set of numerical functions, including addition ($Add$), subtraction ($Sub$), average ($Avg$), limiter ($Lim$), inverse ($Inv$), absolute ($Abs$), multiplication ($Mul$); a set of Boolean functions, including AND gate ($And$), OR gate ($Or$) and negation ($Not$); and also a set of ``combined'' functions, including greater comparison ($Gcom$) and less comparison ($Lcom$). 

Consistent with our notation in the rest of this paper, in this appendix, $\VR{x}$, $\VA{x}$ and $\XF{x}$ depict the reported value, intended value, and the failure mode at variable $\X{x}$, respectively. The same applies to any other variable named here. 

\subsection{Numerical functions}\label{Sec_Basic_NumFun}

\subsubsection{Addition:}\label{Sec_AdditionFMRB}

For $\X{y}, \X{x}_1, \X{x}_2 \in \mathbb{R}$ where $\X{x}_1, \X{x}_2$ are two suspicious variables and $\X{y}\Defs Add(\X{x}_1, \X{x}_2) \Defs\X{x}_1 + \X{x}_2$, the certain causes and minimum conditions for an output failure are as follows:
\begin{eqnarray}
\{\XF{x}_1=\XF{y} \lor \XF{x}_2=\XF{y}\}
&~~Add~~& 
\{\XF{y}=\VF{h} \vee \XF{y}=\VF{l}\} \label{Eq_Add_Proof0}
\end{eqnarray}

\textit{Proof:} Let $\delta \in \mathbb{R}^+$ be a positive real-valued numbers. We first show that $\XF{x}_1=\VF{l}$ is a 
certain cause for  $\XF{y}=\VF{l}$, as follows:
\begin{align}
(\XF{x}_1=\VF{l} \wedge \XF{x}_2=\VF{m})  & \Imply (\VR{x}_1 = \VA{x}_1 - \delta \wedge \VR{x}_2 = \VA{x}_2) \nonumber\\
& \Imply \VR{y} = \VA{y} - \delta \nonumber\\
& \Imply \XF{y}=\VF{l} 
\end{align}

Similarly, we can show that $\XF{x}_2=\VF{l}$ and $(\XF{x}_1=\VF{l} \wedge \XF{x}_2=\VF{l})$ are two certain causes. Thus, we will have the following:
\begin{align}
\XF{y}=\VF{l} &\Imply (\XF{x}_1=\VF{l} \vee \XF{x}_2=\VF{l}) \label{Eq_Add_Proof1}
\end{align}

In the same way, it can be shown that $\XF{x}_1=\VF{h}$, $\XF{x}_2=\VF{h}$, and $(\XF{x}_1=\VF{h} \wedge \XF{x}_2=\VF{h})$ are the certain causes for $\XF{y}=\VF{h}$. Therefore:
\begin{align}
\XF{y}=\VF{h} &\Imply (\XF{x}_1=\VF{h} \vee \XF{x}_2=\VF{h}) \label{Eq_Add_Proof2}
\end{align}
which together with \Eq{Eq_Add_Proof1} means:
\begin{align}
(\XF{y}=\VF{l} \vee \XF{y}=\VF{h}) &\Imply (\XF{x}_1=\XF{y} \lor \XF{x}_2=\XF{y})
\end{align}
which is the logical representation of the proposed model in \Eq{Eq_Add_Proof0}. 

Where $\X{x}_2$ is a certain variable, $\XF{x}_2=\VF{l}$ and $\XF{x}_2=\VF{h}$ are ruled out, and the model in \Eq{Eq_Add_Proof0} will be reduced to:
\begin{eqnarray}
\{\XF{x}_1=\XF{y}\}
&~~Add~~& 
\{\XF{y}=\VF{h} \vee \XF{y}=\VF{l}\} 
\end{eqnarray}

Note that $Add$ is an increasing function with respect to both $\X{x}_1$ and $\X{x}_2$. Therefore, we could as well use \Corol{Corollary_Mon3} to prove \Eq{Eq_Add_Proof0}.

\subsubsection{Subtraction:}

For $\X{y}, \X{x}_1, \X{x}_2 \in \mathbb{R}$ where $\X{x}_1, \X{x}_2$ are two suspicious variables and $\X{y}\Defs Sub(\X{x}_1, \X{x}_2) \Defs\X{x}_1 - \X{x}_2$, the certain conditions and minimum conditions for an output failure are as follows:
\begin{eqnarray}
\{\XF{x}_1=\XF{y} \lor \XF{x}_2=\NF\XF{y}\}
&~~Sub~~& 
\{\XF{y}=\VF{h} \vee \XF{y}=\VF{l}\} \label{Eq_Sub_Proof0}
\end{eqnarray}

\textit{Proof:} $Sub$ is an increasing function with respect to $\X{x}_1$ and a decreasing function with respect to $\X{x}_2$. Therefore, as per \Corol{Corollary_Mon3}, we can say the following:
\begin{align}
(\XF{y}=\VF{l} \vee \XF{y}=\VF{h}) &\Imply (\XF{x}_1=\XF{y} \lor \XF{x}_2=\NF\XF{y})
\end{align}
which is the logical representation of the proposed model in \Eq{Eq_Sub_Proof0}. Alternatively, we could prove the failure model \Eq{Eq_Sub_Proof0} by using $\delta \in \mathbb{R}^+$, as we did earlier for the $And$ function.

Obviously, if $\X{x}_1$ is certain and $\X{x}_2$ suspicious, the model \Eq{Eq_Sub_Proof0} will be reduced to the following:
\begin{eqnarray}
\{\XF{x}_2=\NF\XF{y}\}
&~~Sub~~& 
\{\XF{y}=\VF{h} \vee \XF{y}=\VF{l}\} 
\end{eqnarray}
and if $\X{x}_1$ is the suspicious variable and $\X{x}_2$ the certain one, the model \Eq{Eq_Sub_Proof0} will be reduced to:
\begin{eqnarray}
\{\XF{x}_1=\XF{y}\}
&~~Sub~~& 
\{\XF{y}=\VF{h} \vee \XF{y}=\VF{l}\} 
\end{eqnarray}

\subsubsection{Average:}

For $\X{y}, \X{x}_1, \X{x}_2 \in \mathbb{R}$ where $\X{x}_1, \X{x}_2$ are two suspicious variables and $\X{y}\Defs \\ Avg(\X{x}_1, \X{x}_2) \Defs (\X{x}_1 + \X{x}_2)/\V{2}$, the certain causes and minimum conditions for an output failure are as follows:
\begin{eqnarray}
\{\XF{x}_1=\XF{y} \lor \XF{x}_2=\XF{y}\}
&~~Avg~~& 
\{\XF{y}=\VF{h} \vee \XF{y}=\VF{l}\} \label{Eq_Avg_Proof0}
\end{eqnarray}

\textit{Proof:} $Avg$ is an increasing function with respect to both $\X{x}_1$ and $\X{x}_2$. Therefore, as per \Corol{Corollary_Mon3}, we have the following:
\begin{align}
(\XF{y}=\VF{l} \vee \XF{y}=\VF{h}) &\Imply (\XF{x}_1=\XF{y} \lor \XF{x}_2=\XF{y})
\end{align}
which is the logical representation of the proposed model in \Eq{Eq_Avg_Proof0}. Readers can also refer to \cite{Ref_230_1} for an alternative proof, using failure truth tables.

\subsubsection{Limiter:}

A limiter function is one that limits its real-valued argument between a lower and an upper boundary. Let $\X{x}, \X{p}_l, \X{p}_u \in \mathbb{R}$ be three variables. We define $\X{y}$ as a limiter function, as follows:
\begin{align}
&\X{y}\Defs Lim(\X{x}, \X{p}_l, \X{p}_u) \Defs
\begin{cases}
  	\X{p}_u & \text{if $\X{x} \geq \X{p}_u$}\\
  	\X{x} & \text{$\X{p}_l \leq \X{x} \leq \X{p}_u$}\\
  	\X{p}_l & \text{if $\X{x} \leq \X{p}_l$}\\
\end{cases}
\end{align}
Our proposed model for this function and for a suspicious $\X{x}$ and certain $\X{p}_l$ and $\X{p}_u$ is as follows:
\begin{eqnarray}
\{ \XF{x}=\XF{y}\}
&~~Lim~~& 
\{\XF{y}=\VF{h} \vee \XF{y}=\VF{l}\}
\end{eqnarray}

\textit{Proof:} The $Lim$ function defined above is an increasing function with respect to $\X{x}$. Therefore, as per \Corol{Corollary_Mon3}, the proposed model is correct.

\subsubsection{Inverse:}

Let $\X{x} \in \mathbb{R}$ be a suspicious variable and $\X{y} \Defs Inv(\X{x}) \Defs 1 / \X{x}$. Our proposed model is as follows:
\begin{eqnarray}
\{ \XF{x}=\NF \XF{y}\}
&~~Inv~~& 
\{\XF{y}=\VF{h} \vee \XF{y}=\VF{l}\}
\end{eqnarray}

\textit{Proof:} The $Inv$ function defined above is a decreasing function with respect to $\X{x}$. Therefore, as per \Corol{Corollary_Mon3}, the proposed model is correct.

\subsubsection{Absolute value:}

Let $\X{x} \in \mathbb{R}$ be a suspicious variable and $\X{y} \Defs Abs(\X{x}) \Defs |\X{x}|=sgn(\X{x})\X{x}$. It is not possible to determine a certain cause in $Abs$ function. Deviation in either direction $\VF{l}$ or $\VF{h}$ at the input may result in either direction at the output, depending on the scale of the deviation. Suppose the correct value is $\VA{x}=\V{5}$. A deviation to $\VR{x}=\V{1}$, which is a $\XF{x}=\VF{l}$, causes a $\XF{y}=\VF{l}$ at the output, but a deviation to $\VR{x}=\V{-6}$, although it is a $\XF{x}=\VF{l}$ too, leads to $\XF{y}=\VF{h}$. Therefore, we can only define the minimum conditions, as follows:
\begin{eqnarray}
\{ \XF{x}=\VF{h} \vee \XF{x}=\VF{l}\}
&~~Abs~~& 
\{\XF{y}=\VF{h} \vee \XF{y}=\VF{l}\}
\end{eqnarray}

This model can only be narrowed down if we have additional information about the sign of $\X{x}$. Let $\X{s}=sgn(\X{x})$ and $\X{a}=|\X{x}|$; thus, $\X{x}=\X{s}\X{a}$. If we know that $\XF{s}=\VF{m}$ during the occurrence of the failure, then we can say $\XF{x}=\Sign{\X{s}}\XF{a}=\Sign{\X{x}}\XF{a}$. On the other hand, $\X{y}$ is an increasing function with respect to $\X{a}$. Therefore: 
\begin{align}
((\XF{y}=\VF{l} \vee \XF{y}=\VF{h}) \wedge \XF{s}=\VF{m}) &\Imply \XF{a}=\XF{y} \nonumber \\
&\Imply \XF{x}=\Sign{\X{x}}\XF{y} 
\end{align}
which can be summarized in:
\begin{eqnarray}
\{ \XF{x}=\Sign{\X{x}}\XF{y} \}
&~~Abs~~& 
\{(\XF{y}=\VF{l} \vee \XF{y}=\VF{h}) \wedge \XF{s}=\VF{m})\} \label{Eq_Abs1}
\end{eqnarray}
Note that the right-hand side of \Eq{Eq_Abs1} expresses the premises of the reasoning, part of which is the effect $(\XF{y}=\VF{l} \vee \XF{y}=\VF{h})$.

\subsubsection{Multiplication:}
For $\X{y}, \X{x}, \X{p} \in \mathbb{R}$ where $\X{x}$ is suspicious and $\X{p}$ certain and $\X{y}\Defs Mul\\(\X{x}, \X{p})\Defs\X{x} \times \X{p}$:

\begin{eqnarray}
\{ \XF{x}=\Sign{\X{p}}\XF{y} \}
&~~Mul~~& 
\{\XF{y}=\VF{h} \vee \XF{y}=\VF{l}\} \label{Eq_Mul_1}
\end{eqnarray}

\textit{Proof for \Eq{Eq_Mul_1}:} Follows from the definition of the sign function $\Sign{\X{x}}$.

However, defining the failure modes of $Mul$ for two suspicious variables is rather complicated. Let  $\X{x}_1, \X{x}_2 \in \mathbb{R}$ be two suspicious variables and $\X{y}\Defs Mul(\X{x}_1, \X{x}_2)\\\Defs\X{x}_1 \times  \X{x}_2$. In a value-independent context, the effect of the signs of  $\X{x}_1$ and $\X{x}_2$ prevents us from nominating specific certain causes. An increase (or decrease) in either input may result in either an increase or a decrease at the output. Thus, the minimum conditions for $Mul$ can be defined as follows:
\begin{eqnarray}
\{\XF{x}_1=\VF{h} \lor \XF{x}_1=\VF{l} \lor \XF{x}_2=\VF{h} \lor \XF{x}_2=\VF{l}\}
&~~Mul~~& 
\{\XF{y}=\VF{h} \vee \XF{y}=\VF{l}\} \label{Eq_Mul_2}
\end{eqnarray}

The above statement is very broad. It simply suggests that if output $\X{y}$ is faulty, then at least one of the inputs $\X{x}_1$ and $\X{x}_2$ must be faulty. This outcome is still helpful in some scenarios in which we analyze the input faults and would like to examine all possibilities of failure. However, we can narrow this model further down if we know more about the input variables. One interesting case is when we know that only one input is faulty. In this scenario, the following model can be used:
\begin{eqnarray}
\{ \XF{x}_1=\Sign{\X{x}_2}\XF{y} \vee \XF{x}_2=\Sign{\X{x}_1}\XF{y}\}
&~Mul~& 
\{(\XF{y}=\VF{h} \vee \XF{y}=\VF{l}) \wedge (\XF{x}_1=\VF{m} \vee \XF{x}_2=\VF{m})\}\qquad\qquad \label{Eq_Mul_3}
\end{eqnarray}
which is basically an extension of \Eq{Eq_Mul_1}. 


In another scenario, suppose we know the signs of $\X{x}_1$ and $\X{x}_2$, and we know that the signs will not change as part of the fault. This is obviously a context-based assumption, but not an unrealistic one; in real systems, we often do know the upper and lower bounds for the variables. Let $\X{s}_1=sgn(\X{x}_1)$ and $\X{s}_2=sgn(\X{x}_2)$ represent the signs of  $\X{x}_1$ and $\X{x}_2$. The following model can be used for $Mul$:
\begin{eqnarray}
\{ \XF{x}_1=\Sign{\X{x}_2}\XF{y} \vee \XF{x}_2=\Sign{\X{x}_1}\XF{y}\}
&~Mul~& 
\{(\XF{y}=\VF{h} \vee \XF{y}=\VF{l}) \wedge (\XF{s}_1=\VF{m} \wedge \XF{s}_2=\VF{m})\}\qquad\qquad \label{Eq_Mul_4}
\end{eqnarray}

The term $(\XF{s}_1=\VF{m} \wedge \XF{s}_2=\VF{m})$ indicates that we know that the signs of the two inputs do not change during the occurrence of failure. Note that the conclusion parts in both \Eq{Eq_Mul_3} and \Eq{Eq_Mul_4} are the same, but the premises are different.

\textit{Proof for \Eq{Eq_Mul_4}:} Let $\X{s}_1=sgn(\X{x}_1),\X{s}_2=sgn(\X{x}_2),\X{s}=sgn(\X{y})$, and $\X{a}_1=|\X{x}_1|, \X{a}_2=|\X{x}_2|, \X{a}=|\X{y}|$; thus, $\X{x}_1=\X{s}_1\X{a}_1, \X{x}_2=\X{s}_2\X{a}_2$, and $\X{y}=\X{s}\X{a}$. Given that $\X{a}_1, \X{a}_2 \in \mathbb{R}^+$, $\X{a}$ is an increasing function with respect to both $\X{a}_1$ and $\X{a}_2$. Thus:
\begin{align}
(\XF{a}=\VF{h} \vee \XF{a}=\VF{l}) &\Imply (\XF{a}_1=\XF{a} \vee \XF{a}_2=\XF{a}) \nonumber \\
(\Sign{\X{s}}\XF{y}=\VF{h} \vee \Sign{\X{s}}\XF{y}=\VF{l}) &\Imply (\Sign{\X{s}_1}\XF{x}_1=\Sign{\X{s}}\XF{y} \vee \Sign{\X{s}_2}\XF{x}_2=\Sign{\X{s}}\XF{y}) \nonumber \\
(\XF{y}=\VF{h} \vee \XF{y}=\VF{l}) &\Imply (\Sign{\X{s}_1}\XF{x}_1=\Sign{\X{s}_1}\Sign{\X{s}_2}\XF{y} \vee \Sign{\X{s}_2}\XF{x}_2=\Sign{\X{s}_1}\Sign{\X{s}_2}\XF{y}) \nonumber \\
(\XF{y}=\VF{h} \vee \XF{y}=\VF{l}) &\Imply (\XF{x}_1=\Sign{\X{s}_2}\XF{y} \vee \XF{x}_2=\Sign{\X{s}_1}\XF{y}) \nonumber \\
(\XF{y}=\VF{h} \vee \XF{y}=\VF{l}) &\Imply (\XF{x}_1=\Sign{\X{x}_2}\XF{y} \vee \XF{x}_2=\Sign{\X{x}_1}\XF{y})
\end{align}
which is the logical expression of the model in \Eq{Eq_Mul_4}.

At a different level, where possible (e.g., in system testing scenarios), we can examine the reported values of the variables to draw conclusions. Using truth tables to group failure scenarios, an alternative model for $Mul$ can be as follows:     
\begin{eqnarray}
\{ \XF{x}_1=\VF{h} \lor \XF{x}_2=\VF{h} \lor (\XF{x}_1=\VF{l} \wedge \XF{x}_2=\VF{l}) \}
&~Mul~& 
\{\XF{y}=\VF{l} \wedge \VR{x}_1<0 \wedge \VR{x}_2<0 \}\qquad\qquad\\
\{ \XF{x}_2=\VF{h} \lor (\XF{x}_1=\VF{l} \wedge \XF{x}_2=\VF{l}) \}
&~Mul~& 
\{\XF{y}=\VF{l} \wedge \VR{x}_1<0 \wedge \VR{x}_2=0 \}\\ 
\{ \XF{x}_1=\VF{l} \lor \XF{x}_2=\VF{h} \lor \VA{x}_1=\V{0} \lor \VA{x}_2=\V{0} \}
&~Mul~& 
\{\XF{y}=\VF{l} \wedge \VR{x}_1<0 \wedge \VR{x}_2>0 \}\\ 
\{ \XF{x}_1=\VF{h} \lor (\XF{x}_1=\VF{l} \wedge \XF{x}_2=\VF{l}) \}
&~Mul~& 
\{\XF{y}=\VF{l} \wedge \VR{x}_1=0 \wedge \VR{x}_2<0 \}\\
\{ (\XF{x}_1=\VF{h} \wedge \XF{x}_2=\VF{h}) \lor (\XF{x}_1=\VF{l} \wedge \XF{x}_2=\VF{l}) \}
&~Mul~& 
\{ \XF{y}=\VF{l} \wedge \VR{x}_1=0 \wedge \VR{x}_2=0 \}\\
\{ \XF{x}_1=\VF{l} \lor (\XF{x}_1=\VF{h} \wedge \XF{x}_2=\VF{h}) \}
&~Mul~& 
\{ \XF{y}=\VF{l} \wedge \VR{x}_1=0 \wedge \VR{x}_2>0 \}\\
\{ \XF{x}_1=\VF{h} \lor \XF{x}_2=\VF{l} \lor \VA{x}_1=\V{0} \lor \VA{x}_2=\V{0} \}
&~Mul~& 
\{ \XF{y}=\VF{l} \wedge \VR{x}_1>0 \wedge \VR{x}_2<0 \}\\
\{ \XF{x}_2=\VF{l} \lor (\XF{x}_1=\VF{h} \wedge \XF{x}_2=\VF{h}) \}
&~Mul~& 
\{ \XF{y}=\VF{l} \wedge \VR{x}_1>0 \wedge \VR{x}_2=0 \} \\
\{ \XF{x}_1=\VF{l} \lor \XF{x}_2=\VF{l} \lor (\XF{x}_1=\VF{h} \wedge \XF{x}_2=\VF{h}) \}
&~Mul~& 
\{ \XF{y}=\VF{l} \wedge \VR{x}_1>0 \wedge \VR{x}_2>0 \} \\
\{ \XF{x}_1=\VF{l} \lor \XF{x}_2=\VF{l} \lor \VA{x}_1=\V{0} \lor \VA{x}_2=\V{0} \}
&~Mul~& 
\{ \XF{y}=\VF{h} \wedge \VR{x}_1<0 \wedge \VR{x}_2<0 \}\\ 
\{ \XF{x}_2=\VF{l} \lor (\XF{x}_1=\VF{l} \wedge \XF{x}_2=\VF{h}) \}
&~Mul~& 
\{ \XF{y}=\VF{h} \wedge \VR{x}_1<0 \wedge \VR{x}_2=0 \}\\ 
\{ \XF{x}_1=\VF{h} \lor \XF{x}_2=\VF{l} \lor (\XF{x}_1=\VF{l} \wedge \XF{x}_2=\VF{h}) \}
&~Mul~& 
\{ \XF{y}=\VF{h} \wedge \VR{x}_1<0 \wedge \VR{x}_2>0 \}\\
\{ \XF{x}_1=\VF{l} \lor (\XF{x}_1=\VF{h} \wedge \XF{x}_2=\VF{l}) \}
&~Mul~& 
\{ \XF{y}=\VF{h} \wedge \VR{x}_1=0 \wedge \VR{x}_2<0 \} \\
\{ (\XF{x}_1=\VF{h} \wedge \XF{x}_2=\VF{l}) \lor (\XF{x}_1=\VF{l} \wedge \XF{x}_2=\VF{h}) \}
&~Mul~& 
\{ \XF{y}=\VF{h} \wedge \VR{x}_1=0 \wedge \VR{x}_2=0 \} \\
\{ \XF{x}_1=\VF{h} \lor (\XF{x}_1=\VF{l} \wedge \XF{x}_2=\VF{h}) \}
&~Mul~& 
\{ \XF{y}=\VF{h} \wedge \VR{x}_1=0 \wedge \VR{x}_2>0 \} \\
\{ \XF{x}_1=\VF{l} \lor \XF{x}_2=\VF{h} \lor (\XF{x}_1=\VF{h} \wedge \XF{x}_2=\VF{l}) \}
&~Mul~& 
\{ \XF{y}=\VF{h} \wedge \VR{x}_1>0 \wedge \VR{x}_2<0 \} \qquad\qquad\\
\{ \XF{x}_2=\VF{h} \lor (\XF{x}_1=\VF{h} \wedge \XF{x}_2=\VF{l}) \}
&~Mul~& 
\{ \XF{y}=\VF{h} \wedge \VR{x}_1>0 \wedge \VR{x}_2=0 \}\\ 
\{ \XF{x}_1=\VF{h} \lor \XF{x}_2=\VF{h} \lor \VA{x}_1=\V{0} \lor \VA{x}_2=\V{0} \}
&~Mul~& 
\{ \XF{y}=\VF{h} \wedge \VR{x}_1>0 \wedge \VR{x}_2>0 \}
\end{eqnarray}

\subsection{Boolean functions}\label{Sec_Basic_BoolFun}

\subsubsection{AND gate:}

For $\X{y}, \X{x}_1, \X{x}_2 \in \mathbb{B}$, where $\X{x}_1, \X{x}_2$ are two suspicious variables and $\X{y}\Defs And(\X{x}_1, \X{x}_2) \Defs\X{x}_1 \wedge \X{x}_2$, we propose failure models at three levels. As the minimum conditions, we use:
\begin{eqnarray}
\{ \XF{x}_1=\XF{y} \vee \XF{x}_2=\XF{y}\} &~~And~~& \{\XF{y}=\VF{t} \vee \XF{y}=\VF{f}\}\label{Eq_And_001}
\end{eqnarray}
For certain causes, based on \Def{Def_Certain_Cause}, we use the following:
\begin{eqnarray}
\{ \XF{x}_1=\XF{y} \wedge \XF{x}_2=\XF{y}\} &~~And~~& \{\XF{y}=\VF{t} \} \label{Eq_And_002}\\
\{ \XF{x}_1=\XF{y} \vee \XF{x}_2=\XF{y}\} &~~And~~& \{ \XF{y}=\VF{f} \} \label{Eq_And_003}
\end{eqnarray}
And for value-dependent modeling, we use the following:
\begin{eqnarray}
\{ (\XF{x}_1=\XF{y} \wedge \VR{x}_2=\True) \vee (\XF{x}_2=\XF{y} \wedge \VR{x}_1=\True)\} &~~And~~& \{\XF{y}=\VF{t}\} \label{Eq_And_004}\\
\{ (\XF{x}_1=\XF{y} \wedge \VA{x}_2=\True) \vee (\XF{x}_2=\XF{y} \wedge \VA{x}_1=\True)\} &~~And~~& \{\XF{y}=\VF{f}\} \label{Eq_And_005}
\end{eqnarray}

\textit{Proof:} Let us begin with value-dependent scenarios \Eq{Eq_And_004} and \Eq{Eq_And_005}. From the definition of $And$, it follows that:
\begin{align}
\XF{y}=\VF{f} &\Imply (\VR{y}=\False \wedge \VA{y}=\True) \nonumber \\
&\Imply ((\VR{x}_1=\False \vee \VR{x}_2=\False) \wedge (\VA{x}_1=\True \wedge \VA{x}_2=\True)) \nonumber \\
&\Imply ((\XF{x}_1=\VF{f} \wedge \VA{x}_2=\True) \vee (\XF{x}_2=\VF{f} \wedge \VA{x}_1=\True)) \label{Eq_AND_01} \\
\XF{y}=\VF{t} &\Imply (\VR{y}=\True \wedge \VA{y}=\False) \nonumber \\
&\Imply ((\VR{x}_1=\True \wedge \VR{x}_2=\True) \wedge (\VA{x}_1=\False \vee \VA{x}_2=\False)) \nonumber \\
&\Imply ((\XF{x}_1=\VF{t} \wedge \VR{x}_2=\True) \vee (\XF{x}_2=\VF{t} \wedge \VR{x}_1=\True)) \label{Eq_AND_02} 
\end{align}

Statements \Eq{Eq_AND_01} and \Eq{Eq_AND_02} logically represent the failure scenarios given in \Eq{Eq_And_004} and \Eq{Eq_And_005}.

The minimum conditions for $\XF{y}=\VF{f}$ and $\XF{y}=\VF{t}$ simply follow from \Eq{Eq_AND_01} and \Eq{Eq_AND_02}, as follows:
\begin{align}
\XF{y}=\VF{f} &\Imply (\XF{x}_1=\VF{f} \vee \XF{x}_2=\VF{f}) \label{Eq_AND_03} \\
\XF{y}=\VF{t} &\Imply (\XF{x}_1=\VF{t} \vee \XF{x}_2=\VF{t}) \label{Eq_AND_04}
\end{align}
which, together, represent the failure scenarios given in \Eq{Eq_And_001}.

To draw  the certain conditions, we use truth tables. \Table{Table_AND_TruthTable} shows the failure states related to all the possible combinations of inputs, and their corresponding output failure modes.

Let us examine $\XF{y}=\VF{f}$ first, which represents a deviation between $\VA{y}=\True$ and $\VR{y}=\False$. We claim that $\XF{x}_1=\VF{f}$ is a certain cause for this output failure. According to \Def{Def_Certain_Cause}, we should look for those rows in \Table{Table_AND_TruthTable} that indicate $\VA{y}=\True$, $\XF{x}_1=\VF{f}$, and $\XF{x}_2=\VF{m}$, which includes Row 8 only. Since $\X{x}_2$ is unchanged in this scenario, the only cause of $\XF{y}=\VF{f}$ can be $\XF{x}_1=\VF{f}$. Similarly, we can show that $\XF{x}_2=\VF{f}$ and the conjunction $(\XF{x}_1=\VF{f} \wedge \XF{x}_2=\VF{f})$ are two other certain causes; see Rows 14 and 6, respectively. Therefore, all the certain causes for $\XF{y}=\VF{f}$ can be summarized in: $(\XF{x}_1=\VF{f} \vee \XF{x}_2=\VF{f})$.

The situation for $\XF{y}=\VF{t}$ is somewhat different. From \Eq{Eq_AND_04}, we know that the certain cause for this failure, if any, would be related to $\XF{x}_1=\VF{t}$ and $\XF{x}_2=\VF{t}$. Let us try to see if we can claim $\XF{x}_1=\VF{t}$ as a certain cause. As we did earlier, we need to examine those rows in \Table{Table_AND_TruthTable} where $\VA{y}=\False$, $\XF{x}_1=\VF{t}$, and $\XF{x}_2=\VF{m}$, which include Rows 9 and 12. However, the output failures in these two rows are not consistent; therefore, we cannot claim that $\XF{x}_1=\VF{t}$ always causes $\XF{y}=\VF{t}$. Similarly, we cannot claim that $\XF{x}_2=\VF{t}$ is a certain cause for $\XF{y}=\VF{t}$ (Rows 3 and 15). We can only claim $(\XF{x}_1=\VF{t} \wedge \XF{x}_2=\VF{t})$ as a certain cause for $\XF{y}=\VF{t}$ (see Row 11).

In summary, the certain causes of failure in the $And$ function can be expressed by the following:
\begin{align}
\XF{y}=\VF{f} &\Imply (\XF{x}_1=\VF{f} \vee \XF{x}_2=\VF{f}) \label{Eq_AND_05} \\
\XF{y}=\VF{t} &\Imply (\XF{x}_1=\VF{t} \wedge \XF{x}_2=\VF{t}) \label{Eq_AND_06}
\end{align}
which represent the failure scenarios given in \Eq{Eq_And_002} and \Eq{Eq_And_003}.

Note: both the minimum conditions (given in \Eq{Eq_AND_03} and \Eq{Eq_AND_04}) and the certain causes (given in \Eq{Eq_AND_05} and \Eq{Eq_AND_06}) are logically correct; however, these two outcomes present two different approaches to reasoning: the former is obviously easier to obtain but provides more disjunctive outcomes, whereas the latter requires more effort to reason but provides more conjunctive terms. For some functions (e.g., $Not$), the minimum conditions and certain causes are the same. 

\begin{table}[!h] 
\begin{center}
\begin{tabular}{|M{0.5cm}|M{1.0cm}|M{1.0cm}|M{1.0cm}|M{1.0cm}|M{1.0cm}|M{1.0cm}|M{1.0cm}|M{1.0cm}|M{1.0cm}|}
\hline \multicolumn{1}{|c|}{\TE{No.}} & \multicolumn{1}{c|}{$\VR{x}_1$} & \multicolumn{1}{c|}{$\VA{x}_1$} & \multicolumn{1}{c|}{$\XF{x}_1$} & \multicolumn{1}{c|}{$\VR{x}_2$} & \multicolumn{1}{c|}{$\VA{x}_2$} & \multicolumn{1}{c|}{$\XF{x}_2$} & \multicolumn{1}{c|}{$\VR{y}$} & \multicolumn{1}{c|}{$\VA{y}$} & \multicolumn{1}{c|}{$\XF{y}$}\\ 
\specialrule{0.2em}{0.0em}{0.0em} 
\TE{1} & $\False$ & $\False$ & $\VF{m}$ & $\False$ & $\False$ & $\VF{m}$ & $\False$ & $\False$ & $\VF{m}$ \\ & & & & & & & & & \\[-2em] & & & & & & & & & \\[-0.2em]
\TE{2} & $\False$ & $\False$ & $\VF{m}$ & $\False$ & $\True$ & $\VF{f}$ & $\False$ & $\False$ & $\VF{m}$ \\ & & & & & & & & & \\[-2em] & & & & & & & & & \\[-0.2em]
\TE{3} & $\False$ & $\False$ & $\VF{m}$ & $\True$ & $\False$ & $\VF{t}$ & $\False$ & $\False$ & $\VF{m}$ \\ & & & & & & & & & \\[-2em] & & & & & & & & & \\[-0.2em]
\TE{4} & $\False$ & $\False$ & $\VF{m}$ & $\True$ & $\True$ & $\VF{m}$ & $\False$ & $\False$ & $\VF{m}$ \\ & & & & & & & & & \\[-2em] & & & & & & & & & \\[-0.2em]
\TE{5} & $\False$ & $\True$ & $\VF{f}$ & $\False$ & $\False$ & $\VF{m}$ & $\False$ & $\False$ & $\VF{m}$ \\ & & & & & & & & & \\[-2em] & & & & & & & & & \\[-0.2em]
\TE{6} & $\False$ & $\True$ & $\VF{f}$ & $\False$ & $\True$ & $\VF{f}$ & $\False$ & $\True$ & $\VF{f}$ \\ & & & & & & & & & \\[-2em] & & & & & & & & & \\[-0.2em]
\TE{7} & $\False$ & $\True$ & $\VF{f}$ & $\True$ & $\False$ & $\VF{t}$ & $\False$ & $\False$ & $\VF{m}$ \\ & & & & & & & & & \\[-2em] & & & & & & & & & \\[-0.2em]
\TE{8} & $\False$ & $\True$ & $\VF{f}$ & $\True$ & $\True$ & $\VF{m}$ & $\False$ & $\True$ & $\VF{f}$ \\ & & & & & & & & & \\[-2em] & & & & & & & & & \\[-0.2em]
\TE{9} & $\True$ & $\False$ & $\VF{t}$ & $\False$ & $\False$ & $\VF{m}$ & $\False$ & $\False$ & $\VF{m}$ \\ & & & & & & & & & \\[-2em] & & & & & & & & & \\[-0.2em]
\TE{10} & $\True$ & $\False$ & $\VF{t}$ & $\False$ & $\True$ & $\VF{f}$ & $\False$ & $\False$ & $\VF{m}$ \\ & & & & & & & & & \\[-2em] & & & & & & & & & \\[-0.2em]
\TE{11} & $\True$ & $\False$ & $\VF{t}$ & $\True$ & $\False$ & $\VF{t}$ & $\True$ & $\False$ & $\VF{t}$ \\ & & & & & & & & & \\[-2em] & & & & & & & & & \\[-0.2em]
\TE{12} & $\True$ & $\False$ & $\VF{t}$ & $\True$ & $\True$ & $\VF{m}$ & $\True$ & $\False$ & $\VF{t}$ \\ & & & & & & & & & \\[-2em] & & & & & & & & & \\[-0.2em]
\TE{13} & $\True$ & $\True$ & $\VF{m}$ & $\False$ & $\False$ & $\VF{m}$ & $\False$ & $\False$ & $\VF{m}$ \\ & & & & & & & & & \\[-2em] & & & & & & & & & \\[-0.2em]
\TE{14} & $\True$ & $\True$ & $\VF{m}$ & $\False$ & $\True$ & $\VF{f}$ & $\False$ & $\True$ & $\VF{f}$ \\ & & & & & & & & & \\[-2em] & & & & & & & & & \\[-0.2em]
\TE{15} & $\True$ & $\True$ & $\VF{m}$ & $\True$ & $\False$ & $\VF{t}$ & $\True$ & $\False$ & $\VF{t}$ \\ & & & & & & & & & \\[-2em] & & & & & & & & & \\[-0.2em]
\TE{16} & $\True$ & $\True$ & $\VF{m}$ & $\True$ & $\True$ & $\VF{m}$ & $\True$ & $\True$ & $\VF{m}$ \\
[0.3em]\hline 
\multicolumn{10}{|c|}{
$\VR{y}= \VR{x}_1 \wedge \VR{x}_2$, $\VA{y}= \VA{x}_1 \wedge \VA{x}_2$}\\
\multicolumn{10}{|c|}{
$\XF{x}_1=\MD(\VR{x}_1, \VA{x}_1)$, $\XF{x}_2=\MD(\VR{x}_2, \VA{x}_2)$, $\XF{y}=\MD(\VR{y}, \VA{y})$
}\\
[0.3em]\hline 
\end{tabular}
\end{center}
\caption{Failure truth table for $And$}\label{Table_AND_TruthTable}
\end{table}

\subsubsection{OR gate:}

For $\X{y}, \X{x}_1, \X{x}_2 \in \mathbb{B}$, where $\X{x}_1, \X{x}_2$ are two suspicious variables and $\X{y}\Defs Or(\X{x}_1, \X{x}_2) \Defs\X{x}_1 \vee \X{x}_2$, we propose failure models at three levels. As the minimum conditions, we use the following:
\begin{eqnarray}
\{ \XF{x}_1=\XF{y} \vee \XF{x}_2=\XF{y}\} &~~Or~~& \{\XF{y}=\VF{t} \vee \XF{y}=\VF{l}\} \label{Eq_Or_Pos_Model}
\end{eqnarray}

For certain causes, based on \Def{Def_Certain_Cause}, we use the following:
\begin{eqnarray}
\{ \XF{x}_1=\XF{y} \vee \XF{x}_2=\XF{y}\} &~~Or~~& \{\XF{y}=\VF{t} \} \label{Eq_Or_CC_Model1}\\
\{ \XF{x}_1=\XF{y} \wedge \XF{x}_2=\XF{y}\} &~~Or~~& \{ \XF{y}=\VF{f} \} \label{Eq_Or_CC_Model2}
\end{eqnarray}
and for value-dependent modeling, we use the following:
\begin{eqnarray}
\{ (\XF{x}_1=\XF{y} \wedge \VA{x}_2=\False) \vee (\XF{x}_2=\XF{y} \wedge \VA{x}_1=\False)\} &~~Or~~& \{\XF{y}=\VF{t} \} \label{Eq_Or_VB_Model1}\\
\{ (\XF{x}_1=\XF{y} \wedge \VR{x}_2=\False) \vee (\XF{x}_2=\XF{y} \wedge \VR{x}_1=\False)\} &~~Or~~&  \{ \XF{y}=\VF{f} \} \label{Eq_Or_VB_Model2}
\end{eqnarray}

\textit{Proof:} From $\XF{y}=\VF{f}$ and $\XF{y}=\VF{t}$ we will have:
\begin{align}
\XF{y}=\VF{f} &\Imply (\VR{y}=\False \wedge \VA{y}=\True) \nonumber \\
&\Imply ((\VR{x}_1=\False \wedge \VR{x}_2=\False) \wedge (\VA{x}_1=\True \vee \VA{x}_2=\True)) \nonumber \\
&\Imply ((\XF{x}_1=\VF{f} \wedge \VR{x}_2=\False) \vee (\XF{x}_2=\VF{f} \wedge \VR{x}_1=\False)) \label{Eq_OR_01_01}\\
\XF{y}=\VF{t} &\Imply (\VR{y}=\True \wedge \VA{y}=\False) \nonumber \\
&\Imply ((\VR{x}_1=\True \vee \VR{x}_2=\True) \wedge (\VA{x}_1=\False \wedge \VA{x}_2=\False)) \nonumber \\
&\Imply ((\XF{x}_1=\VF{t} \wedge \VA{x}_2=\False) \vee (\XF{x}_2=\VF{t} \wedge \VA{x}_1=\False)) \label{Eq_OR_02_01} 
\end{align}

\Eq{Eq_OR_01_01} and \Eq{Eq_OR_02_01} prove the value-dependent models in \Eq{Eq_Or_VB_Model1} and \Eq{Eq_Or_VB_Model2}. From \Eq{Eq_OR_01_01} and \Eq{Eq_OR_02_01}, we can also conclude the following:
\begin{align}
\XF{y}=\VF{f} &\Imply (\XF{x}_1=\VF{f} \vee \XF{x}_2=\VF{f}) \label{Eq_OR_03_01} \\
\XF{y}=\VF{t} &\Imply (\XF{x}_1=\VF{t} \vee \XF{x}_2=\VF{t}) \label{Eq_OR_04_01} 
\end{align}
which together prove \Eq{Eq_Or_Pos_Model}. 

To demonstrate the correctness of \Eq{Eq_Or_CC_Model1} and \Eq{Eq_Or_CC_Model2}, we will use truth tables. \Table{Table_OR_TruthTable} is the failure truth table for an $Or$ function. 

\begin{table}[!h]
\begin{center}
\begin{tabular}{|M{0.5cm}|M{1.0cm}|M{1.0cm}|M{1.0cm}|M{1.0cm}|M{1.0cm}|M{1.0cm}|M{1.0cm}|M{1.0cm}|M{1.0cm}|N|}
\hline \multicolumn{1}{|c|}{\TE{No.}} & \multicolumn{1}{c|}{$\VR{x}_1$} & \multicolumn{1}{c|}{$\VA{x}_1$} & \multicolumn{1}{c|}{$\XF{x}_1$} & \multicolumn{1}{c|}{$\VR{x}_2$} & \multicolumn{1}{c|}{$\VA{x}_2$} & \multicolumn{1}{c|}{$\XF{x}_2$} & \multicolumn{1}{c|}{$\VR{y}$} & \multicolumn{1}{c|}{$\VA{y}$} & \multicolumn{1}{c|}{$\XF{y}$}\\ 
\specialrule{0.2em}{0.0em}{0.0em} 
\TE{1} & $\False$ & $\False$ & $\VF{m}$ & $\False$ & $\False$ & $\VF{m}$ & $\False$ & $\False$ & $\VF{m}$ \\ & & & & & & & & & \\[-2em] & & & & & & & & & \\[-0.2em]
\TE{2} & $\False$ & $\False$ & $\VF{m}$ & $\False$ & $\True$ & $\VF{f}$ & $\False$ & $\True$ & $\VF{f}$ \\ & & & & & & & & & \\[-2em] & & & & & & & & & \\[-0.2em]
\TE{3} & $\False$ & $\False$ & $\VF{m}$ & $\True$ & $\False$ & $\VF{t}$ & $\True$ & $\False$ & $\VF{t}$ \\ & & & & & & & & & \\[-2em] & & & & & & & & & \\[-0.2em]
\TE{4} & $\False$ & $\False$ & $\VF{m}$ & $\True$ & $\True$ & $\VF{m}$ & $\True$ & $\True$ & $\VF{m}$ \\ & & & & & & & & & \\[-2em] & & & & & & & & & \\[-0.2em]
\TE{5} & $\False$ & $\True$ & $\VF{f}$ & $\False$ & $\False$ & $\VF{m}$ & $\False$ & $\True$ & $\VF{f}$ \\ & & & & & & & & & \\[-2em] & & & & & & & & & \\[-0.2em]
\TE{6} & $\False$ & $\True$ & $\VF{f}$ & $\False$ & $\True$ & $\VF{f}$ & $\False$ & $\True$ & $\VF{f}$ \\ & & & & & & & & & \\[-2em] & & & & & & & & & \\[-0.2em]
\TE{7} & $\False$ & $\True$ & $\VF{f}$ & $\True$ & $\False$ & $\VF{t}$ & $\True$ & $\True$ & $\VF{m}$ \\ & & & & & & & & & \\[-2em] & & & & & & & & & \\[-0.2em]
\TE{8} & $\False$ & $\True$ & $\VF{f}$ & $\True$ & $\True$ & $\VF{m}$ & $\True$ & $\True$ & $\VF{m}$ \\ & & & & & & & & & \\[-2em] & & & & & & & & & \\[-0.2em]
\TE{9} & $\True$ & $\False$ & $\VF{t}$ & $\False$ & $\False$ & $\VF{m}$ & $\True$ & $\False$ & $\VF{t}$ \\ & & & & & & & & & \\[-2em] & & & & & & & & & \\[-0.2em]
\TE{10} & $\True$ & $\False$ & $\VF{t}$ & $\False$ & $\True$ & $\VF{f}$ & $\True$ & $\True$ & $\VF{m}$ \\ & & & & & & & & & \\[-2em] & & & & & & & & & \\[-0.2em]
\TE{11} & $\True$ & $\False$ & $\VF{t}$ & $\True$ & $\False$ & $\VF{t}$ & $\True$ & $\False$ & $\VF{t}$ \\ & & & & & & & & & \\[-2em] & & & & & & & & & \\[-0.2em]
\TE{12} & $\True$ & $\False$ & $\VF{t}$ & $\True$ & $\True$ & $\VF{m}$ & $\True$ & $\True$ & $\VF{m}$ \\ & & & & & & & & & \\[-2em] & & & & & & & & & \\[-0.2em]
\TE{13} & $\True$ & $\True$ & $\VF{m}$ & $\False$ & $\False$ & $\VF{m}$ & $\True$ & $\True$ & $\VF{m}$ \\ & & & & & & & & & \\[-2em] & & & & & & & & & \\[-0.2em]
\TE{14} & $\True$ & $\True$ & $\VF{m}$ & $\False$ & $\True$ & $\VF{f}$ & $\True$ & $\True$ & $\VF{m}$ \\ & & & & & & & & & \\[-2em] & & & & & & & & & \\[-0.2em]
\TE{15} & $\True$ & $\True$ & $\VF{m}$ & $\True$ & $\False$ & $\VF{t}$ & $\True$ & $\True$ & $\VF{m}$ \\ & & & & & & & & & \\[-2em] & & & & & & & & & \\[-0.2em]
\TE{16} & $\True$ & $\True$ & $\VF{m}$ & $\True$ & $\True$ & $\VF{m}$ & $\True$ & $\True$ & $\VF{m}$ \\
[0.3em]\hline 
\multicolumn{10}{|c|}{
$\VR{y}= \VR{x}_1 \vee \VR{x}_2$, $\VA{y}= \VA{x}_1 \vee \VA{x}_2$}\\
\multicolumn{10}{|c|}{
$\XF{x}_1=\MD(\VR{x}_1, \VA{x}_1)$, $\XF{x}_2=\MD(\VR{x}_2, \VA{x}_2)$, $\XF{y}=\MD(\VR{y}, \VA{y})$.
}\\
[0.3em]\hline 
\end{tabular}
\end{center}
\caption{Failure truth table for $Or$}\label{Table_OR_TruthTable}
\end{table}

From \Eq{Eq_OR_04_01}, we know that the certain cause for this failure is related to $\XF{x}_1=\VF{t}$ and $\XF{x}_2=\VF{t}$. For $\XF{x}_1=\VF{t}$ as a certain cause, we need to examine those rows in \Table{Table_OR_TruthTable} where $\VA{y}=\False$, $\XF{x}_1=\VF{t}$, and $\XF{x}_2=\VF{m}$, which only includes Row 9. Since $\X{x}_2$ is unchanged in this scenario, the only cause of $\XF{y}=\VF{t}$ can be $\XF{x}_1=\VF{t}$. Similarly, we can show that $\XF{x}_2=\VF{t}$ and the conjunction $\XF{x}_1=\VF{t} \wedge \XF{x}_2=\VF{t}$ are two other certain causes (see Rows 3 and 11, respectively). For $\XF{y}=\VF{f}$, the only certain cause will be in Row 6. The output failure is not consistent in the other rows with only one of the two input failures  $\XF{x}_1=\VF{f}$ and $\XF{x}_2=\VF{f}$. In summary, the certain causes of failure in $Or$ function can be as follows:
\begin{align}
\XF{y}=\VF{f} &\Imply (\XF{x}_1=\VF{f} \wedge \XF{x}_2=\VF{f}) \label{Eq_OR_05_01} \\
\XF{y}=\VF{t} &\Imply (\XF{x}_1=\VF{t} \vee \XF{x}_2=\VF{t}) \label{Eq_OR_06_01}
\end{align}
which represent the logical form of the model in \Eq{Eq_Or_CC_Model1} and \Eq{Eq_Or_CC_Model2}. 

\subsubsection{Boolean negation:}

For $\X{y}, \X{x} \in \mathbb{B}$, where $\X{x}$ is a suspicious variable and $\X{y}\Defs Not(\X{x})\\ \Defs\neg \X{x}$:
\begin{eqnarray}
\{ \XF{x}=\NF \XF{y}\} &~~Not~~& \{\XF{y}=\VF{a}\}
\end{eqnarray}

\textit{Proof:} Follows from the definition of $\neg$ and $\NF$.

\subsection{Combined functions}\label{Sec_Basic_CombFun}

\subsubsection{Greater comparison:}

Let $\X{x}_1, \X{x}_2 \in \mathbb{R}$ and $\X{y} \in \mathbb{B}$ and $\X{y}\Defs Gcom(\X{x}_1, \X{x}_2)\Defs (\X{x}_1>\X{x}_2)$. Where both $\X{x}_1$ and $\X{x}_2$ are suspicious, the minimum conditions for a failure at $\X{y}$ are modeled as follows:
\begin{eqnarray}
\{ \XF{x}_1= \VF{h} \vee \XF{x}_2= \VF{l}\}
&~~Gcom~~& 
\{\XF{y}=\VF{t}\} \label{Eq_Gcom_001}\\
\{ \XF{x}_1= \VF{l} \vee \XF{x}_2= \VF{h}\}
&~~Gcom~~& 
\{\XF{y}=\VF{f}\} \label{Eq_Gcom_002}
\end{eqnarray}
If either $\X{x}_1$ or $\X{x}_2$ is certain, its corresponding sentence can be removed from the above disjunctions, and the other variable becomes the certain cause.

\textit{Proof:} The correctness of minimum conditions \Eq{Eq_Gcom_001} and \Eq{Eq_Gcom_002} follows from the definition of $Gcom$: if the output $\X{y}$ is faulty, it can only due to a fault at $\X{x}_1$ or $\X{x}_2$, and only in the fault directions given here in \Eq{Eq_Gcom_001} and \Eq{Eq_Gcom_002}. It is also obvious that if either input is certain, then the only cause of the fault could be the other input and, again, in the direction of fault indicated in \Eq{Eq_Gcom_001} and \Eq{Eq_Gcom_002}. Readers can also refer to \cite{Ref_230_1} for a proof of single suspicious variable scenarios based on failure truth table.

\subsubsection{Less comparison:}

Let $\X{x}_1, \X{x}_2 \in \mathbb{R}$ and $\X{y} \in \mathbb{B}$ and $\X{y}\Defs Lcom(\X{x}_1, \X{x}_2)\Defs (\X{x}_1<\X{x}_2)$. Where both $\X{x}_1$ and $\X{x}_2$ are suspicious, the minimum conditions for a failure at $\X{y}$ are modeled as follows:
\begin{eqnarray}
\{ \XF{x}_1= \VF{l} \vee \XF{x}_2= \VF{h}\}
&~~Lcom~~& 
\{\XF{y}=\VF{t}\}\\
\{ \XF{x}_1= \VF{h} \vee \XF{x}_2= \VF{l}\}
&~~Lcom~~& 
\{\XF{y}=\VF{f}\}
\end{eqnarray}
If either $\X{x}_1$ or $\X{x}_2$ is certain, its corresponding sentence can be removed from the above disjunctions, and the other variable becomes the certain cause.

\textit{Proof:} Similar to $Gcom$.

\end{document}